\documentclass[acmsmall]{acmart}

\makeatletter
\def\@acmplainindent{0pt}
\def\@acmdefinitionindent{0pt}
\def\@proofindent{\noindent}
\makeatother

\usepackage{local}

\newboolean{isExtendedVersion}
\setboolean{isExtendedVersion}{true} 

\newcommand{\appendixref}{
    \ifthenelse{\boolean{isExtendedVersion}}%
    {the appendix}
    {the extended version of this paper {\cite{Moeller2025Extended}}}
}

\setcopyright{cc}
\setcctype{by}
\acmDOI{10.1145/3729295}
\acmYear{2025}
\acmJournal{PACMPL}
\acmVolume{9}
\acmNumber{PLDI}
\acmArticle{192}
\acmMonth{6}
\received{2024-11-15}
\received[accepted]{2025-03-06}



\begin{document}

\title{Active Learning of Symbolic \NetKAT Automata}

\author{Mark Moeller}
\orcid{0009-0002-9512-565X}
\affiliation{%
  \institution{Cornell University}
  \city{Ithaca}
  \country{USA}
}
\email{moeller@cs.cornell.edu}

\author{Tiago Ferreira}
\orcid{0000-0002-6942-0228}
\affiliation{%
  \institution{University College London}
  \city{London}
  \country{United Kingdom}
}
\email{t.ferreira@ucl.ac.uk}

\author{Thomas Lu}
\orcid{0009-0000-3474-6263}
\affiliation{%
  \institution{Cornell University}
  \city{Ithaca}
  \country{USA}
}
\email{cl2625@cornell.edu}

\author{Nate Foster}
\orcid{0000-0002-6557-684X}
\affiliation{%
  \institution{Cornell University}
  \city{Ithaca}
  \country{USA}
}
\affiliation{%
  \institution{Jane Street}
  \city{New York}
  \country{USA}
}
\email{jnfoster@cs.cornell.edu}

\author{Alexandra Silva}
\orcid{0000-0001-5014-9784}
\affiliation{%
  \institution{Cornell University}
  \city{Ithaca}
  \country{USA}
}
\email{alexandra.silva@cornell.edu}

\begin{abstract}
NetKAT is a domain-specific programming language and logic that has been successfully used to specify and verify the behavior of packet-switched networks. This paper develops techniques for automatically learning \NetKAT models of unknown networks using active learning. Prior work has explored active learning for a wide range of automata (e.g., deterministic, register, B\"{u}chi, timed etc.) and also developed applications, such as validating implementations of network protocols. We present algorithms for learning different types of \NetKAT automata, including symbolic automata proposed in recent work. We prove the soundness of these algorithms, build a prototype implementation, and evaluate it on a standard benchmark. Our results highlight the applicability of symbolic \NetKAT learning for realistic network configurations and topologies.
\end{abstract}

\titlenote{This research was developed with funding from the Defense Advanced Research Projects Agency (DARPA) and the Office of Naval Research (ONR). The views, opinions, and/or findings contained in this material are those of the authors and should not be interpreted as representing the official views or policies of the Department of the Navy, DARPA, or the U.S. Government. The U.S. Government is authorized to reproduce and distribute reprints for Government purposes notwithstanding any copyright notation hereon. DISTRIBUTION A. Approved for public release; distribution unlimited.}

\maketitle

\section{Introduction}

Model-checking is one of the success stories in formal methods, with broad applications in hardware and software. However, model checking requires having accurate models of the systems being verified. These models are usually produced by human experts via a slow and error-prone process.

Active learning is an appealing alternative that can construct an accurate model of a system automatically. Model learning is often used with closed-box systems---i.e., when the source code or inner workings of a system cannot be inspected, but its external behaviors can be observed---and it generates formal models, such as finite automata, that can be used to obtain assurance through testing or verification. In particular, Angluin's \lstar algorithm, which learns a Deterministic Finite Automaton (DFA) in the so-called Minimally Adequate Teacher (MAT) framework, has inspired nearly 40 years of fruitful research~\cite{kearnsvazirani1994,Bollig,bonakdarpour_ttt_2014,Cassel2014,Zetzsche2022}.

The \lstar approach is most applicable when only observations of behavior are available and it is reasonable to assume the system has a finite state model, or can be approximated by such a model. Computer networking is one such domain, which has both of these characteristics. We frequently have only partial information about the policies used to route packets through the network, but we do have the ability to observe packet-level behaviors using command-line tools like \texttt{ping} and \texttt{traceroute}. At the same time, the programs that execute on network routers are finite state---their behavior is kept intentionally simple to enable them to operate at line rate. Indeed, automata learning has been applied to learning implementations of network protocols \cite{Brostean2014,Ferreira2021}.

In this paper, we want to go a step further and learn not just end-host protocols, but network-wide behaviors. This would be helpful in many scenarios: inferring a model of a single device where the program and configuration are unavailable, to verify that it conforms to a well-understood model; inferring the topology of the network and check that it conforms to the expected structure; or inferring a model an unknown peer network to verify that it is fulfilling a customer agreement.

Indeed, the networking community has recognized that there is great value in having accurate models of individual devices and the network as a whole. For example, Google employs a small team of verification engineers who develop and maintain models of their data center switches in a domain-specific language \cite{Albab2022}. They use these models to find bugs (using fuzzers) and verify properties (using symbolic execution tools). However, maintaining the models turns out to be non-trivial---switch pipelines are moderately complex, and their behavior changes as vendors add new features. Using automata learning, rather than having to expend effort developing these models by hand, they could be generated automatically. Learned models also have the advantage of being \emph{ground-truth} in the sense that they are based on actual observations of system behavior.

This paper presents a framework for automated learning of \NetKAT automata. To achieve this goal, we bring together two lines of research: active learning and algebraic approaches to network verification. We develop new automata learning algorithms for the expressive models used in \NetKAT, a domain-specific programming language and logic based on Kleene Algebra with Tests \cite{Anderson2014}.  \NetKAT allows us to model different aspects of a network at different levels of detail and abstraction, applicable to all the scenarios described above. Moreover, as \NetKAT was designed with an automata-theoretic foundation, it is a great fit for active learning. The key challenges, however, are in designing specialized data structures and algorithms that take advantage of \NetKAT's algebraic structure and other details of the networking domain, to scale up to real-world networks.

In particular, while \NetKAT has an elegant theory based on using packets as actions in the automaton model, for learning network policies, it has an obvious drawback: the space of packets is exponential in the number of header bits---intractably large for all but toy policies. To overcome the large packet space challenge, recent work introduced a symbolic form of \NetKAT automata~\cite{Moeller2024}. Crucially, these  automata are symbolic in {\em both} the transition labels and the state space. Hence, these symbolic \NetKAT automata are therefore the ideal target for exploring learning.

Symbolic \NetKAT automata are reminiscent of Symbolic Finite Automata (SFAs)~\cite{symbolic-automata}, a fundamental model introduced to deal with infinite or intractably large alphabets in automata. The key idea in SFAs is that rather than transitions being explicitly defined by the singular element of the alphabet they consume, one can have transitions be labeled by predicates that determine the subset of the alphabet that can take a specific transition. SFAs have enjoyed a range of useful applications~\cite{hutchison_applications_2013}, largely due to their simple theory and efficient decision procedures, and extensions of classic automata learning algorithms have been proposed for subclasses of symbolic automata~\cite{symbolic-learning,argyros2018,fisman_inferring_2023}. Of particular relevance is recent work on learning a large subclass of SFAs~\cite{argyros2018}.

At first glance, prior work on learning SFAs seems to provide the recipe for dealing with \NetKAT automata: just compile the \NetKAT automata down to DFAs, and use SFA techniques to deal with the large alphabet. Unfortunately, there are several key differences that make  this idea a nonstarter: the semantics of \NetKAT automata allow transitions to be based on a notion of a \emph{carry-on packet} (or current packet), which means that the naive translation of \NetKAT automata to DFAs also triggers a blow-up of \emph{state}. This is precisely the issue addressed in symbolic \NetKAT automata~\cite{Moeller2024}, and is the reason why we cannot re-use learning algorithms for SFAs with \NetKAT.

This paper develops active learning for \NetKAT automata, making the following contributions:

\begin{itemize}[-,leftmargin=*]
\item \textbf{An expanded theory of \NetKAT automata}: Angluin-style learning algorithms crucially rely on the classic Myhill-Nerode theorem, which identifies the states of the minimal automaton with equivalence classes of languages. We present a new Myhill-Nerode theorem for \NetKAT (\Cref{sec:myhillnerode}) that enable us to characterize a \emph{canonical} form for \NetKAT automata.

\item \textbf{A naive learning algorithm}: We give a ``first attempt'' solution for learning canonical \NetKAT automata, and prove its correctness (\Cref{sec:learningcanonical}). It is naive in that it does not use symbolic techniques, and it therefore has poor performance due to the large ``alphabet'' and state space.

\item \textbf{Two algorithms for symbolic \NetKAT learning}: To provide scalable learning for \NetKAT, we also develop an algorithm for learning symbolic \NetKAT automata. We do this in two stages. First, we focus on the subset of \NetKAT that does not have the $\dup$ primitive (i.e., the $\dup$-free fragment). We give an algorithm for learning symbolic representations of \NetKAT programs in this restricted setting (\Cref{sec:learningspp}). Next, using this first symbolic learner as a subroutine for symbolic learning of full-blown \NetKAT automata and prove its correctness (\Cref{sec:learningsymbolic}).

\item \textbf{A prototype OCaml implementation}: We have implemented our symbolic learning algorithms as a prototype to investigate their behavior and properties. We conclude our paper, by describing our implementation and experience learning models for network policies from a standard benchmark suite (\Cref{sec:evaluation}).
\end{itemize}
Omitted proofs can be found in\appendixref.

\section{Background on \NetKAT and \NetKAT Automata}
\label{sec:netkat}

\begin{figure}[t]
    \renewcommand{\arraystretch}{1.1}
    \begin{tabular}{lcc}
        \textbf{Syntax} & \textbf{Description} & \textbf{Semantics} $\Sem{p} \subseteq  \Pk \cdot (\Pk \cdot \dup)^\star \cdot \Pk$\\[1mm]
        \hline \\[-0.9em]
        $p,q \Coloneqq\ \ ~\bot$ & \emph{False} & $\emptyset$ \\
        \qquad$\mid\quad \top$ & \emph{True} & $\{ \alpha\alpha \mid \alpha\in\Pk \}$ \\
        \qquad$\mid\quad f \test v$ & \emph{Test equals} & $\{ \alpha\alpha \mid \alpha.f = v \}$ \\
        \qquad$\mid\quad f \testNE v$ & \emph{Test not equals} & $\{ \alpha\alpha \mid \alpha.f \neq v \}$ \\
        \qquad$\mid\quad f \mut\, v$ & \emph{Modification} & $\{ \alpha\alpha[f \mut\, v] \mid \alpha\in\Pk \}$ \\
        \qquad$\mid\quad \text{dup}$ & \emph{Duplication} & $\{ \alpha\cdot (\alpha\cdot \dup)\cdot\alpha \mid \alpha\in\Pk \}$ \\
        \qquad$\mid\quad p + q$ & \emph{Union} & $\Sem{p} \cup \Sem{q}$ \\
        \qquad$\mid\quad p \cdot q$ & \emph{Sequencing} & $\Sem{p}\diamond \Sem{q}$ \\
        \qquad$\mid\quad p^\star$ & \emph{Iteration} & $\bigcup\limits_{n\geq 0} \Sem{p^n}$ s.t. $p^0 \triangleq \one$, $p^{n+1} \triangleq p\cdot p^n$
        \\[3mm]
        \hline \\[-2.8em]
\phantom{        {Values} $v \Coloneqq 0 \mid \ldots $ }&
\phantom{            {Fields} $f \Coloneqq f_1 \mid \ldots \mid f_k$ }&
\phantom{           {Packets} $\pk \Coloneqq \{f_1 = v_1, \ldots, f_k = v_k\}$}
    \end{tabular}
    \renewcommand{\arraystretch}{1.0}
    \caption{\NetKAT syntax and semantics.}
    \label{fig:synsem}
\end{figure}

\NetKAT~\cite{Anderson2014} was introduced as a language for reasoning about network routing policies with axiomatic, decidable program equivalence as a design principle.  This section briefly reviews \NetKAT and \NetKAT automata as developed in prior work.  We fix a set of packet header fields $F = \{f_1, \ldots, f_n\}$, and a finite set of header values $V$. We write records (finite maps) as a collection of mappings. For example, a \emph{packet} is a finite record assigning values to fields $\pk = \{f_1 \mapsto v_1, \ldots, f_n \mapsto v_n\}$. We reference the field $f_1$ of packet $\pk$ by $\pk.f_1=v_1$. The set of all packets is denoted $\Pk = F \Mapsto V$.

The syntax and semantics of \NetKAT are presented in \Cref{fig:synsem}. The basic primitives in \NetKAT are packet tests ($f \test v$, $f \testNE v$) and modifications ($f\mut v$). Program expressions are then compositionally built from tests and packet modifications, using union ($+$), sequencing ($\cdot$), and iteration ($\star$).  Conditionals and loops can be encoded in the standard way:
\[
\textbf{if } b\textbf{ then } p  \textbf{ else } q \triangleq b \cdot p + \neg b \cdot q \qquad\qquad
\textbf{while } b\textbf{ do } p   \triangleq (b \cdot p)^\star \cdot \neg b.
\]
In a network, conditionals can be used to model the behavior of the forwarding tables on individual switches while iteration can be used to model the iterated processing performed by the network as a whole; the original paper on \NetKAT provides further details~\cite{Anderson2014}. Note that assignments and tests in \NetKAT are always against constant values. This is a key restriction that makes equivalence decidable and aligns the language with the capabilities of network hardware. The $\text{dup}$ primitive makes a copy of the current packet and appends it to the trace, which only ever grows as the packet goes through the network. This primitive is crucial for expressing network-wide properties, as it allows the semantics to ``observe''  intermediate packets at internal switches.

The semantics of \NetKAT is based on regular sets of {\em guarded strings} which are elements of the set $\GS = { \Pk \cdot (\Pk \cdot \dup)^\star \cdot \Pk}$. Here, $\dup$ can be thought of as delimiting ``letters'' in the string. Equivalently, we can think of guarded strings as elements of the isomorphic set $\Pk \cdot \Pk^\star \cdot \Pk$ which are strings over the alphabet $\Pk$ with at least two letters. Given two guarded strings $\alpha x \beta$ and $ \gamma y \xi $ their concatenation, denoted by $\diamond$, is defined only when $\beta = \gamma$:
\begin{align}\label{def:gsconc}
\alpha x \beta \diamond \gamma y \xi = \begin{cases} \alpha x y \xi  & \beta = \gamma\\
\text{undefined} &  \beta \neq \gamma\\
\end{cases}
\end{align}
This definition lifts to sets of guarded strings $U,V \subseteq \GS$ pointwise: $U \diamond V = \{ u \diamond v \mid u\in U\text{ and }v\in V\}$.

\NetKAT's formal semantics, given in \Cref{fig:synsem} associates sets of guarded strings to each expression.  The semantics for $\top$ gives all traces containing the input/output packet $\alpha\alpha$, whereas $\bot$ produces no traces.  Tests ($f \test v$ and $f \testNE v$) produce filter traces, depending on whether the test succeeds or fails. Modifications ($f \mut v$) produce a trace with the modified packet.  Duplication ($\text{dup}$) produces traces with two copies of the input packet $\pk$.  Union ($p + q$) produces the union of the traces produced by $p$ and $q$.  Sequential composition ($p \cdot q$)
produces the concatenation of the traces produced by $p$ and $q$, where the last packet in output traces of $p$ is used as the input to $q$. Finally, iteration ($p^\star$) produces the union of the traces produced by concatenation of $p$, iterated zero or more times.

\subsection{Encoding Networks in \NetKAT}\label{sec:encoding}

\begin{wrapfigure}[4]{R}{0.25\textwidth}
\vspace{-15pt}
\begin{tikzpicture}
    \node[draw, circle] (A) at (0,0) {1};
    \node[draw, circle] (B) at (2,0) {2};
    \draw (A) -- node[above,pos=0.2] {2} node[above,pos=0.8] {2} (B);
    \draw (A) --++ (-0.5,0.5) node[midway, above left] {1};
    \draw (A) --++ (-0.5,-0.5) node[midway, below left] {3};
    \draw (B) --++ (0.5,0.5) node[midway, above right] {3};
    \draw (B) --++ (0.5,-0.5) node[midway, below right] {1};
\end{tikzpicture}
\end{wrapfigure}

We encode the behavior of networks in \NetKAT using a similar approach as in prior work \cite{Anderson2014,Foster2015,Smolka2015,Moeller2024}. For example, suppose we have a toy network with two switches with three ports each, connected by a link between both switches' port 2:

We use two special fields $\sw$ and $\pt$, which denote a packet's location at a particular switch and port. Using these fields, we can encode the network topology in \NetKAT---e.g., the expression,
\[\textsf{t} \triangleq \pt\test 1 + \pt\test 3 + \pt\test 2\cdot(\sw\test 1\cdot \sw\mut 2 + \sw\test 2\cdot\sw\mut 1)\]

Similarly, we can encode routing policies for individual switches in the network above:
\[r \triangleq \pt\test 1\cdot \pt\mut 2 + \pt\test 2\cdot\pt\mut 1,\]
which encodes the policy: ``At either switch, forward packets arriving on port 1 via port 2, and forward packets arriving on port 2 via port 1.'' This policy drops packets arriving at port 3.

Using these encodings, we can compose the topology ($t$) and routing policy ($r$) into one expression capturing the network's \emph{transfer function}: $r\cdot t$. We combine this with ingress and egress predicates $p_i$ and $p_f$ to encode the full end-to-end behavior:
$p_i \cdot (r \cdot t \cdot \dup)^\star \cdot p_f$.
Here we use a $\dup$ to denote that we want to log one copy of each packet for each network ``hop'' (i.e. an update to $\sw$ in $t$). The star indicates that we are interested in all paths allowed by this policy on this topology. The expressions $p_i$ and $p_f$ are predicates which constrain the initial and final packets we are interested in. For example, to investigate all paths by which Switch 1 (Port 1) can reach Switch 2 (Port 1), we might choose
$p_i \triangleq \sw\test 1 \cdot \pt\test 1$, and
$p_f \triangleq \sw\test 2 \cdot \pt\test 1$.
More complicated routing policies may involve other pertinent fields, perhaps  $\dst$, encoding a destination switch.

It is the aim of this paper to begin to answer the question: Can we build a model that completely describes this network and its policy by sending packets and observing how they are processed? After building up the component processes, we return to this example in \Cref{sec:netex}.

\subsection{\NetKAT Automata}

Despite the fact that assignments and tests in \NetKAT programs are always against constants, the behavior of \NetKAT programs can be complicated, particularly when conditionals, iteration, and multiple packet fields are involved. To provide a more operational view of the semantics, \NetKAT automata were introduced by \citet{Foster2015} as acceptors of guarded strings.
\begin{definition}\label{def:nka}
A \NetKAT automaton (NKA) is a four-tuple $(S,s_0,\delta,\varepsilon)$, where $S$ is a finite set of states, $s_0 \in S$ is the start state, $\delta$ and $\varepsilon$ are the transition and observation functions, respectively\footnote{We use $\bin$ to denote the Booleans, $\{\zero,\one\}$.}:
$$
\delta\colon S \to\Pk\to\Pk\to S\qquad\qquad\varepsilon\colon S \to\Pk\to\Pk \to \bin
$$
We will write $\varepsilon_{\alpha \beta}(s)$ as a shorthand for $\varepsilon(s)(\alpha) (\beta)$ and $\delta_{\alpha \beta}(s)$ as a shorthand for $\delta(s)(\alpha) (\beta)$.
\end{definition}
	Semantically, an $\NKA$ $\mathcal{M} = (S, s_0, \delta, \varepsilon)$ accepts a language $\mathcal{L}(\mathcal{M}) \subseteq \GS$  containing all strings $s \in \GS$ for which $\mathcal{M}(s_0, w) = \top$, where:
	\begin{align}\label{def:semantics}
	\mathcal{M}(s, \pk\cdot\pkp) = \varepsilon_{\alpha\beta} (s) \qquad\qquad
	\mathcal{M}(s, \pk\cdot\pkp\cdot\dup\cdot w') = \mathcal{M}(\delta_{\pk\pkp}(q), \pkp\cdot w')
	\end{align}
Note how in the semantics of state $s$ of \NetKAT automata we read two packets $\alpha\beta$ moving to state $\delta_{\alpha\beta}(s)$ \emph{but} we retain $\beta$ in the string that remains! So, unlike traditional regular expressions and automata, \NetKAT is stateful---in effect, the packet $\beta$ has not been fully consumed, but is used as a {\em carry-on} packet to remember the last step of the automaton's execution. This peculiarity of \NetKAT complicates the semantics (i.e., it is not a straightforward language semantics) and designing equivalence and minimization procedures, which have to account for the carry-on packet. We can fold the packet into the state of the automaton and approach the semantics and the problem of learning using classical automata, but this results in enormous state blowup and is therefore impractical. This is the key challenge for designing a learning algorithm we address in this paper.

\section{Myhill-Nerode Theorem for \NetKAT and Canonical Automata}\label{sec:myhillnerode}

The essential insight of the \lstar algorithm is that we can identify Myhill-Nerode equivalence classes for DFAs in an incremental fashion. However, to use this idea in the context of \NetKAT, we need to develop a \NetKAT-specific notion of Myhill-Nerode. This section develops such a notion, which we will use as a basis for a learning algorithm in \Cref{sec:learningcanonical}. Our approach follows Kozen's more general treatment of Myhill-Nerode for automata on guarded strings \cite{Kozen2001}.

For a language $\mathcal{L}$ over an alphabet $\Sigma$, the
Myhill-Nerode~\cite{Nerode1958} relation for $\mathcal{L}$ is an
equivalence relation on $\Sigma^\star$, written $(\equiv_\mathcal{L})$
and defined by:
\[
s \equiv_\mathcal{L} t\stackrel \triangle  \iff \forall e\in \Sigma^\star. (se \in \mathcal{L} \iff te \in \mathcal{L})
\]
The Myhill-Nerode theorem states that $\mathcal{L}$ is regular iff $\equiv_\mathcal{L}$ has finite index. Moreover, the equivalence classes of $\equiv_\mathcal{L}$ correspond precisely to the states of the {\em unique minimal} DFA for $\mathcal{L}$. We would like an analogous characterization for \NetKAT, but there are two key challenges: concatenation of guarded strings is not always defined (\cref{def:gsconc}) and \NetKAT's semantics process strings by reading two packets at a time--one of which continues as {\em carry-on} packet for the next step (\cref{def:semantics}). These differences result in a subtly different characterization for guarded strings.

We first define the set of guarded string prefixes $\Pref = \Pk\cdot(\Pk\cdot\dup)^\star$ and the set of guarded string suffixes $\Suf = (\Pk\cdot\dup)^\star\cdot\Pk$. We also define a helper function $\textsf{last} \colon \Pref \to \Pk$ and a (partial) concatenation operation $\mathbin{\blacklozenge} \colon \Pref \times \GS \to \GS$ for prefixes with guarded strings by:
\[
\begin{array}{ll}
	\last(\pk) & = \pk \\
	\mathbin{\last}(w \cdot \pk \cdot \dup) & = \pk
\end{array}\qquad\qquad\qquad
	p \mathbin{\blacklozenge} \pk \cdot s = \begin{cases}
		p \cdot s & \last(p) = \pk \\
		\text{undefined} & \text{otherwise}
	\end{cases}
\]

For a language $\mathcal{L}\subseteq\GS$, we define the relation $\NKmne$ for $s,t \in \Pref$ by:
\begin{equation}\label{eqn:nkmne}
s \NKmne t \stackrel\triangle\iff (\forall e \in \GS.\ s \mathbin{\blacklozenge} e \in\mathcal{L} \iff t \mathbin{\blacklozenge} e \in\mathcal{L})
\end{equation}

The relation $\NKmne$ is an equivalence on guarded string prefixes.
\begin{rlemma}{nker}
The relation $\NKmne$ in \Cref{eqn:nkmne} is reflexive, symmetric, and transitive.
\end{rlemma}

The equivalence classes of the classic Myhill-Nerode relation for $\mathcal{L} \subseteq \Sigma^\star$ correspond exactly to the states of the minimal DFA accepting $\mathcal{L}$. However, this result does not immediately transfer to \NetKAT. We need to define a simpler form of automata that hides the carry-on packet in the state---this will enable us to precisely capture the fact that the equivalence classes of $\NKmne$ can only relate prefixes $s$ and $t$ when $\last(s) = \last(t).$\footnote{Note that we interpret (\text{$\iff$}) to hold when both sides are undefined or both defined and equal, but it does \emph{not} hold when only one side is undefined (regardless of the other side).}

\newcommand{\spell}{\mathsf{spell}}
\begin{definition}
A \emph{Packet-state} \NetKAT automaton
(\PNKA) is a four-tuple $(Q, q_0, \partial, \lambda)$, where $Q$ is a finite set of states, $q_0: \Pk \to Q$ is the start state, $\partial\colon Q \to \Pk \to Q$ is the transition function, and $\lambda\colon Q \to\Pk \to 2$ is the observation function.

The functions $\partial$ and $\lambda$ must satisfy the \emph{spelling} property: for any states $q,q' \in Q$ and packets $\pk, \pkp \in \Pk$, if $\partial_\pk(q) = \partial_\pkp(q')$, $q_0(\pk) = q_0(\pkp)$, or $q_0(\pk) = \partial_\pkp(q)$, then $\pk = \pkp$.
\end{definition}

The spelling property ensures we cannot arrive at the same state via two different packets. Formally, there is a function $\spell \colon Q \to \Pk$ that outputs the packet of all transitions into each state.

A \PNKA $\mathcal{M} = (Q, q_0, \partial, \lambda)$ accepts a language $\mathcal{L}(\mathcal{M}) \subseteq \GS$ consisting of all strings $\pk \cdot w \in \GS$ satisfying $\mathcal{M}(q_0(\pk), w) = \top$, where:
\[
\mathcal{M}(q, \pkp) = \lambda_{\pkp}(q) \qquad
\mathcal{M}(q, \pkp \cdot \dup \cdot w') = \mathcal{M}(\partial_\pkp(q), w')
\]

As the name suggests, the distinguishing feature of a \PNKA is that the transition function and observation function operate on a single packet: the other packet is ``hidden in the state.'' Although \PNKA satisfy additional restrictions compared to standard \NetKAT automata, it is important to note that they recognize exactly the same sets of guarded strings.

\begin{rtheorem}{nkaiffpnka}
Let $\mathcal{L}\subseteq\GS$. Then $\mathcal{L}$ is accepted by a $\NKA$ if and only if it is accepted by a $\PNKA$.
\end{rtheorem}

Now that we have established that $\PNKA$ are a suitable representation of $\NKA$ we are ready to state the Myhill-Nerode theorem for $\NetKAT$. This theorem ensures $\PNKA$ have properties that are desirable for automata learning.

\begin{definition}
Given a language $\mathcal{L}$ we define $\mathcal{P}_{\NKmne} = (Q, q_0, \partial, \lambda)$ by:
\[
Q = \{[s]_{\NKmne} \mid s \in \Pref \}\qquad
q_0(\pk) = [\pk]_{\NKmne}\qquad
\partial_{\pk}([s]_{\NKmne}) = [s \cdot \pk \cdot \dup]_{\NKmne}
\qquad
\lambda_{\pk}([s]_{\NKmne}) = s \cdot \pk \in \mathcal L
\]
\end{definition}

\begin{rtheorem}{netkatmn}[Myhill-Nerode for $\NetKAT$]
\label{thm:mn}
For a given language $\mathcal{L} \subseteq \GS$:
\begin{enumerate}
\item $\mathcal{L}$ is regular if and only if $\NKmne$ has finite index.
\item For regular $\mathcal L$, a minimal $\PNKA$ accepting $\mathcal L$ has $|\NKmne|$ states.
\item Any $\PNKA$ accepting $\mathcal{L}$ with $|\NKmne|$ states is isomorphic to $\mathcal{P}_{\NKmne}$.
\end{enumerate}
\end{rtheorem}

\section{Learning Canonical \NetKAT Automata}\label{sec:learningcanonical}

In this section, we develop a learning algorithm for \emph{canonical} \NetKAT automata (i.e., minimal \PNKA). This algorithm is primarily of theoretical interest, as canonical automata will have a large number of states in general. However, the core ideas of this approach will guide us in designing a more practical algorithm in \Cref{sec:learningsymbolic} for learning \emph{symbolic} \NetKAT automata.

Our algorithm resembles \lstar, but we cannot use the same state for different ``carry-on'' packets in a \PNKA. Accommodating this restriction requires some additional structure.

\subsection{The \PNKA Teacher}

Just as in Angluin's MAT, the MAT framework for \NetKAT automata assumes a teacher that can answer two types of queries: membership queries (i.e., whether a given guarded string is part of the target language or not) and equivalence queries (i.e., whether a given hypothesis automaton describes the target language, and returning a counter-example if not).

\begin{definition}[MAT Interface]
The MAT interface for Packet-state \NetKAT automata is as follows:
\[
\memPNKA: \GS \to \bin \qquad \equivPNKA : \PNKA \to 1 + \GS
\]
\end{definition}

The teacher knows the target language, $\mathcal L \subseteq \GS$. We have $\memPNKA(w) = \one$ iff $w\in \mathcal L$, and
$\equivPNKA(\mathcal H) = \one$ if $\mathcal{L}(\mathcal H) = \mathcal L$. Otherwise,
$\equivPNKA(\mathcal H) = w \in \mathcal{L}\oplus \mathcal{L}(\mathcal H)$.
In the rest of this section we develop a naive algorithm for learning a canonical \PNKA
from this type of teacher.

\subsection{The Observation Table}\label{sec:observationtable}

The central abstraction used in our learning algorithm is an \emph{observation table}. This data structure keeps track of the knowledge the learner has acquired, and it provides the basis for constructing hypothesis automata. The shape of the table is similar to the classic observation table used in \lstar, but we need to adjust it to the specific nature of \NetKAT automata. In particular, we organize the observation table as a collection of smaller tables, one for each packet in $\Pk$.

\begin{definition}[Packet Table]
For a packet $\pk \in \Pk$, a packet table is a triple $(S_\pk, E_\pk, T_\pk)$, where:
\begin{itemize}[-]
\item $S_\pk \subseteq \{ p \mid \last(p) = \alpha, p \in \Pref \}$ is a set of \emph{access prefixes}.
\item $E_\pk \subseteq \Suf $ is a set of \emph{distinguishing suffixes}.
\item $T_\pk \colon (S_\pk \cup S'_\pk) \times E_\pk \rightarrow \bin$ is such that $T_\pk(s, e) = \one$ if $s \cdot e \in \mathcal{L}$, and $T_\pk(s,e) = \zero$ otherwise.
\end{itemize}
We maintain $S'_\pk \triangleq \bigcup_{\pkp\in\Pk} S_\pkp \cdot (\pk \cdot \dup)$ as the set of $(\pk \cdot \dup)$ \emph{extensions} of all packet table sets $S_\pkp$.
\end{definition}

\begin{definition}[Observation Table]
An observation table $T$ is an $\pk$-indexed set of Packet Tables, $(S_\pk, E_\pk, T_\pk)$, consisting of one packet table per packet in $\Pk$.
\end{definition}

\begin{remark}
In \lstar, the extension part of the table is usually written $S \cdot \Sigma$, and computes the one letter extension of each string representing a state---i.e., the transitions of an automaton. We require a similar functionality in our case, albeit adjusted to the specific details of packet tables. In particular, for a packet table $(S_\pk, E_\pk, T_\pk)$, the set $S'_\pk$ is maintained to have the $(\pk \cdot \dup)$ extension of every state in every packet table. This allows $T_\pk$ to hold information on every $\pk$ transition between any pair of states.
\end{remark}

\paragraph{Access Prefixes}
Access prefixes describe states of the target automaton---the state corresponding to a given access prefix is the one the automaton would land on after processing that prefix. Prefixes comprising a single packet naturally represent the start state, while prefixes comprising two packets (and one $\dup$) represent the states reached by reading a given packet from the start state, and so on.

\paragraph{Distinguishing Sequences}
Distinguishing sequences, on the other hand, elicit specific behavior from an arbitrary state. Note that because the prefixes come from $\Pref$ and distinguishing sequences from $\Suf$, concatenation is always defined and forms a guarded string. Distinguishing sequences can also be understood as the suffixes $e$ in the Myhill-Nerode relation of \Cref{sec:myhillnerode}. In \lstar, the set of distinguishing suffixes is maintained to be suffix-closed by the algorithm. A similar property holds here, but only for the observation table as a whole, not each individual packet table.

\begin{example}
\label{ex:as-and-ds}
Consider the $\PNKA$ $\mathcal{M}$ below over the packet space $\Pk = \{\{f \mapsto 1\},\{f \mapsto 2\}\}$, which we abbreviate to $\pk$ and $\pkp$, respectively. States from equivalence classes ending in $\pk$ are colored orange and lined, while states from equivalence classes ending in $\pkp$ are colored blue and dotted. The dotted transitions from the start state denote its special type $q_0 \colon \Pk \rightarrow Q$ in the automaton.
\begin{center}
\begin{tikzpicture}[bend angle=30]
		\node[state, initial] (q0) {$q_0$};
		\node[state, right of=q0, node distance=55pt, pattern=crosshatch dots, pattern color=blue!50!white] (q1) {$q_1$};
		\node[state, right of=q1, node distance=55pt, pattern=north east lines, pattern color=orange!90!white] (q3) {$q_3$};
		\node[state, right of=q3, node distance=55pt, pattern=north east lines, pattern color=orange!90!white] (q2) {$q_2$};
		\node[state, right of=q2, node distance=55pt, pattern=crosshatch dots, pattern color=blue!50!white] (q4) {$q_4$};

		\node[draw=none, below of=q1, node distance=27pt] {$\{\pk\}$};
		\node[draw=none, below of=q2, node distance=27pt] {$\varnothing$};
		\node[draw=none, below of=q3, node distance=27pt] {$\{\pkp\}$};
		\node[draw=none, below of=q4, node distance=27pt] {$\varnothing$};

		\draw (q0) edge [densely dotted, above] node{$\pk$} (q1);
		\draw (q0) edge [densely dotted, bend left, above] node[ fill=white, anchor=center, pos=0.5,font=\bfseries]{$\pkp$} (q2);
		\draw (q1) edge [bend left] node[ fill=white, anchor=center, pos=0.5,font=\bfseries]{$\pk$} (q4);
		\draw (q1) edge [bend left, below] node{$\pkp$} (q3);
		\draw (q2) edge [bend right, below] node{$\pk$} (q4);
		\draw (q2) edge [loop left] node{$\pkp$} (q2);
		\draw (q3) edge [bend left] node{$\pk$} (q1);
		\draw (q3) edge [bend right, below] node{$\pkp$} (q2);
		\draw (q4) edge [loop right] node{$\pk$} (q4);
		\draw (q4) edge [bend right] node{$\pkp$} (q2);

		\draw (q1) edge[ematrix] ++ (0, -0.75);
		\draw (q2) edge[ematrix] ++ (0, -0.75);
		\draw (q3) edge[ematrix] ++ (0, -0.75);
		\draw (q4) edge[ematrix] ++ (0, -0.75);
	\end{tikzpicture}
\end{center}
Note that there are infinitely many possible access prefixes and distinguishing sequences, due to the cyclic structure of the automaton. Some access prefixes for $q_1$ are $\pk$ and $\pk \cdot \pkp \cdot \dup \cdot \pk$. An access prefix for $q_3$ is $\pk \cdot \pkp \cdot \dup$. Finally, $\pkp$ is a distinguishing sequence for $q_1$ and $q_3$, as $\mathcal{M}(q_0(\pk), \pkp) = \bot$ but $\mathcal{M}(q_0(\pk), \pkp \cdot \dup \cdot \pkp) = \top$.
\end{example}

\paragraph{Closed and Consistent Observation Tables}
In \lstar, the main loop of the algorithm makes membership queries to construct a closed table, from which
a hypothesis can be made. We define an analogous concept here, also in terms of the $\row$ function, just as in \lstar. Namely, we define a primitive row function with type $\row_\pk : S_\pk \cup S'_\pk \to E_\pk \to \bin$, defined by
$\row_\pk(s)(e) = T_\pk(s, e)$. Clearly all we have done here is to curry $T_\pk$---but the currying is essential! We will partially apply $\row_\pk$ to a given string $s$ and consider the resulting functions $E_\pk \to \bin$. In this way, we distinguish states by viewing their behavior in terms of the suffixes we have seen for each packet.

\begin{definition}[Closed Table]
An observation table $T$ is \emph{closed}, if for each $\pk\in\Pk$ and for each $s'\in S'_\pk$, then $\row_\pk(s') = \row_\pk(s)$ for some $s\in S_\pk$.
\end{definition}

\begin{definition}[Consistent Table]\label{def:pnkaconsistent}
An observation table $T$ is \emph{consistent}, if for each $\pk,\pkp\in\Pk$ and for each $s,s'\in S_\pk$, then $\row_\pk(s) = \row_\pk(s') \Rightarrow \row_\pkp(s\cdot\pkp\cdot\dup) = \row_\pkp(s'\cdot\pkp\cdot\dup)$.
\end{definition}

The closedness property can be checked locally to each packet table (provided that all the required extensions to each prefix by each packet have all already been entered into $T$), while the consistency property must be checked globally, since successors to rows in one table may be in another table.

We use the observation table in the same way as in \lstar: by making membership queries until the table is closed and consistent, ensuring that the construction of the next \PNKA to conjecture is well-defined. In \Cref{fig:closedconsist}, we show how to make the table closed and consistent. The function $\extend$ performs membership queries as needed so that every $T_\pk$ is defined for its required domain.

\subsection{The \PNKA Learning Algorithm}

Our $\NKA$ learning algorithm works by iteratively expanding an observation table, which is then the basis from which a hypothesis is constructed. If the hypothesis is correct then the learner terminates. Otherwise, a counterexample from the equivalence oracle is used to refine the tables. The main routine of the algorithm (\Cref{fig:learncanonical}) follows the classic learning loop of MAT-based learners.
\begin{figure}[H]
\begin{subfigure}{.57\textwidth}
      \centering
     \begin{algorithmic}
       \State $T \gets \{\pk \mapsto (S_\pk=\{\pk\}, E_\pk=\Pk, T_\pk=\{\}) \mid \pk\in\Pk \}$
       \While {$\one$}
           \While {$T$ not closed and consistent}
           	   \State $T \gets \textsf{mkConsistent}(\textsf{mkClosed}(T))$
           \EndWhile
           \State $\mathcal{H} \gets \hypPNKA(T)$
           \State $c \gets \equivPNKA(\mathcal{H})$
           \If {$c = \one$}
               \State \textbf{return} $\mathcal{H}$
           \Else
               \State $T \gets \update(T, c)$
           \EndIf
       \EndWhile
     \end{algorithmic}
     \end{subfigure}
		\hfill
    \begin{subfigure}{.42\textwidth}
      \centering
     \begin{algorithmic}
      \State \textbf{Input: }Table $T$, counterex. $c\in\GS$
             \State \textbf{Output: }$T$ updated in place with $c$.
       \State \ 
       \For {$s \in \Pref$, $e \in \Suf$ such that $c = s \cdot e$}
         \State $S_{\last(s)} \gets S_{\last(s)} \cup \{s\}$
     	 \EndFor
       \State \textbf{return} $\extend(T)$
              \State
       \State
       \State
    \end{algorithmic}
         \end{subfigure}
    \caption{Algorithm for learning canonical automata (left) and subroutine for \textsf{update} (right).}\label{fig:learncanonical}
\end{figure}

\begin{figure}[H]
\begin{subfigure}{.45\textwidth}
      \centering
	\begin{algorithmic}
		\If {$T_\pk$ is not closed}
            \State let $s'\in S'_\pk$ such that \\ \;\;\;\;\;\;\; $\row_\pk(s') \notin \{ \row_\pk(s) \mid s \in S_\pk\}$.
            \State $S_\pk \gets S_\pk \cup \{s'\}$
            \State \textbf{return} $\extend(T)$
        \EndIf
	\end{algorithmic}
\end{subfigure}
		\hfill
    \begin{subfigure}{.5\textwidth}
      \centering
	\begin{algorithmic}
    	\If {$T$ is not consistent}
        	\State let $e\in E_\pkp$ such that for some $s,s'\in S_\pk$, $\row_\pk(s)= \row_\pk(s')$, \\ \;\;\;\;\;\;\; but $T_\pkp(s\cdot\pkp\cdot\dup, e) \neq T_\pkp(s'\cdot\pkp\cdot\dup, e)$.
            \State $E_\pk \gets E_\pk \cup \{\pkp\cdot\dup\cdot e\}$
            \State \textbf{return} $\extend(T)$
        \EndIf
	\end{algorithmic}
	\end{subfigure}
	\caption{Subroutines used in \PNKA learning (\cref{fig:learncanonical}): $\textsf{mkClosed} \colon T \rightarrow T$ (left) and $\textsf{mkConsistent} \colon T \rightarrow T$ (right).}\label{fig:closedconsist}
\end{figure}

For the $\update$ operation, we first break the counterexample string at each possible location. Each break results in a prefix ending in some packet $\pk$, which we add to $S_\pk$.

Finally, constructing the hypothesis once the table is closed and consistent is again similar to \lstar, except that we still need to keep the state space associated with each table separate. To make this clear in the formalism, states themselves are a pair of type $\Pk \times (E_\pk\to\bin)$: a packet $\pk$, and a row vector of $T_\pk$.
Given an observation table $T$, we construct $\hypPNKA(T) \triangleq (Q, q_0, \delta, \varepsilon)$, where:
\begin{align*}
&Q = \bigcup \{(\pk,\row_\pk(s)) \mid s\in S_\pk, \pk \in \Pk\} & \delta(\pk, \row_\pk(s))(\pkp) = (\pkp, \row_\pkp(s\cdot\pkp\cdot\dup))\\
&q_0(\pk) = (\pk, \row_\pk(\pk))
&\varepsilon(\pk, \row_\pk(s))(\pkp) = T_\pk (s, \pkp)
	\end{align*}

\begin{example}
Were we to select the $\PNKA$ of \Cref{ex:as-and-ds} as the target, we would have the following final packet tables and hypothesis:

\begin{tabular}{@{}c@{\qquad}c@{\qquad}c}
\begin{tikzpicture}[baseline]
  \matrix (magic) [matrix of nodes]
  {
	$\pk$ & $\pk$ & $\pkp$ \\ \hline
	$\pk$ & |[pattern=crosshatch dots, pattern color=blue!50!white]|$\top$ & |[pattern=crosshatch dots, pattern color=blue!50!white]|$\bot$ \\
	$\pk\pk$ & |[pattern=crosshatch dots, pattern color=blue!50!white]|$\bot$ & |[pattern=crosshatch dots, pattern color=blue!50!white]|$\bot$ \\ \hline
	$\pk\pk$ & $\bot$ & $\bot$ \\
	$\pk\pk\pk$ & $\bot$ & $\bot$ \\
	$\pkp\pk$ & $\bot$ & $\bot$ \\
	$\pk\pkp\pk$ & $\top$ & $\bot$ \\
};
\draw (magic-1-2.north west)--(magic-7-1.south east);
\end{tikzpicture}
\begin{tikzpicture}[bend angle=30, baseline]
		\node[state, initial, minimum size=0.75cm] (q0) {$q_0$};
		\node[state, right of=q0, minimum size=0.72cm, node distance=50pt, pattern=crosshatch dots, pattern color=blue!50!white] (q1) {$\top\bot$};
		\node[state, right of=q1, minimum size=0.72cm, node distance=50pt, pattern=north east lines, pattern color=orange!90!white] (q3) {$\bot\top$};
		\node[state, right of=q3, minimum size=0.72cm, node distance=50pt, pattern=north east lines, pattern color=orange!90!white] (q2) {$\bot\bot$};
		\node[state, right of=q2, minimum size=0.72cm, node distance=50pt, pattern=crosshatch dots, pattern color=blue!50!white] (q4) {$\bot\bot$};

		\node[draw=none, below of=q1, node distance=27pt] {$\{\pk\}$};
		\node[draw=none, below of=q2, node distance=27pt] {$\varnothing$};
		\node[draw=none, below of=q3, node distance=27pt] {$\{\pkp\}$};
		\node[draw=none, below of=q4, node distance=27pt] {$\varnothing$};

		\draw (q0) edge [densely dotted, above] node{$\pk$} (q1);
		\draw (q0) edge [densely dotted, bend left, above] node{$\pkp$} (q2);
		\draw (q1) edge [bend left] node{$\pk$} (q4);
		\draw (q1) edge [bend left, below] node{$\pkp$} (q3);
		\draw (q2) edge [bend right, below] node{$\pk$} (q4);
		\draw (q2) edge [loop left] node{$\pkp$} (q2);
		\draw (q3) edge [bend left] node{$\pk$} (q1);
		\draw (q3) edge [bend right, below] node{$\pkp$} (q2);
		\draw (q4) edge [loop right] node{$\pk$} (q4);
		\draw (q4) edge [bend right] node{$\pkp$} (q2);

		\draw (q1) edge[ematrix] ++ (0, -0.75);
		\draw (q2) edge[ematrix] ++ (0, -0.75);
		\draw (q3) edge[ematrix] ++ (0, -0.75);
		\draw (q4) edge[ematrix] ++ (0, -0.75);
\end{tikzpicture}
\begin{tikzpicture}[baseline]
  \matrix (magic) [matrix of nodes]
  {
	$\pkp$ & $\pk$ & $\pkp$ \\ \hline
	$\pkp$ & |[pattern=north east lines, pattern color=orange!90!white, minimum height=0.59cm]|$\bot$ & |[pattern=north east lines, pattern color=orange!90!white, minimum height=0.59cm]|$\bot$ \\
	$\pk\pkp$ & |[pattern=north east lines, pattern color=orange!90!white, minimum height=0.59cm]|$\bot$ & |[pattern=north east lines, pattern color=orange!90!white, minimum height=0.59cm]|$\top$ \\ \hline
	$\pk\pkp$ & $\bot$ & $\top$ \\
	$\pk\pk\pkp$ & $\bot$ & $\bot$ \\
	$\pkp\pkp$ & $\bot$ & $\bot$ \\
	$\pk\pkp\pkp$ & $\bot$ & $\bot$ \\
};
\draw (magic-1-2.north west)--(magic-7-1.south east);
\end{tikzpicture}
\end{tabular}

\end{example}

\subsection{Correctness of the Canonical Learner}

Next, we prove that the canonical automaton learner is correct.
We start by showing that the learner produces valid conjectures. For succinctness, we will assume the observation tables used to build hypotheses are closed and consistent, as this follows from the definition of the learning algorithm.

\begin{rlemma}{matchestable}
The hypothesis automaton, $\hypPNKA(T)$, is well defined for any Observation Table $T$.
\end{rlemma}

The next lemma says that the learner only conjectures minimal (i.e., canonical) automata.

\begin{rlemma}{hmin}
\label{lem:hmin}
  For any hypothesis \PNKA $\mathcal{H} = (Q, q_0, \partial, \lambda)$ produced by the learner, and for any two prefixes $s,t\in\Pref$,
we have that $s \equivH t$ iff $s \equivLH t$.
\end{rlemma}

The next lemma says that the Myhill-Nerode relation for the target language always refines the access relation for the learner's conjecture. In other words, the learner can only be wrong by failing to distinguish two states that need to be different.

\begin{rlemma}{lrefine}
\label{lem:lrefine}
Suppose the teacher holds $\mathcal{L}$. Then for each hypothesis \PNKA $\mathcal{H}$ of the learner, we have that $(\NKmne)$ refines $(\equivLH)$ in the sense that for prefixes $s,t\in\Pref$, then $s \NKmne t$ implies $s \equivLH t$.
\end{rlemma}

Additionally, hypotheses remain faithful to the data in the observation table they originate from.

\begin{rtheorem}{agree}
\label{thm:agree}
	Given a closed, consistent table $T$, $\mathcal{H} \triangleq \hypPNKA(T)$ agrees on every cell of the packet tables. That is, $T_\pkpp(\pk \cdot s, e) = \mathcal{H}(q_0(\pk), s \cdot e)$ for any $\pk \cdot s \in S_\pkpp \cup S'_\pkpp$, $e \in E_\pkpp$, and $\pkpp \in \Pk$.
\end{rtheorem}

Finally, each counterexample results in a conjecture $\mathcal{H}$ with at least one more state:

\begin{rlemma}{addstate}
\label{lem:addstate}
Fix the target language $\mathcal{L}$, and suppose a learner conjectures a \PNKA $\mathcal{H}$, and we get a counterexample $c = \equivPNKA(\mathcal{H})$, i.e. $c \in \mathcal{L} \oplus \mathcal{L}(\mathcal{H})$.
Then the next conjecture by the learner, $\mathcal{H'}$, will have at least one additional state compared to $\mathcal{H}$.
\end{rlemma}

\begin{rtheorem}{totalcorrectness}
The algorithm in \Cref{fig:learncanonical} terminates with the correct automaton.
\end{rtheorem}
\begin{proof}
It follows that if the learner terminates then the automaton is correct, as we only terminate upon a successful equivalence query.
Because the Myhill-Nerode relation $(\NKmne)$ for the language $\mathcal{L}$ refines those of the hypotheses (\Cref{lem:lrefine}), then once we conjecture a hypothesis $H$ for which $|\NKmne| = |\equivH|$, we know the automaton must be correct by \Cref{thm:mn}.
By \Cref{lem:addstate}, we make progress toward this bound with every counterexample, completing the proof.
\end{proof}

\section{Learning Symbolic Packet Programs (SPPs)}\label{sec:learningspp}

Before we present our algorithm for learning symbolic \NetKAT automata, we first develop the core techniques needed to handle a restricted case: $\dup$-free \NetKAT programs. We then use the learner we develop in this restricted setting as the central component of a learner that handles full \NetKAT. In addition to its role in the full automaton learner, however, this algorithm can be used directly whenever input-output packet pairs are enough to capture the relevant behavior of a system, such as a single closed-box device, or the single-step transfer function of a network.

\subsection{Background}
\citet{Moeller2024} introduced a set of techniques for reasoning about \NetKAT \emph{symbolically}. Their representation first develops a data structure based on Binary Decision Diagrams (BDDs) for the dup-free fragment of \NetKAT which they call a Symbolic Packet Program (SPP).
\[
p \in \SPP \Coloneqq\ \bot \mid \top \mid
    \SPP(f, \{ \ldots, v_i \mapsto \{ \ldots, w_{ij} \mapsto q_{ij}, \ldots \}, \ldots \}, \{ \ldots, w_i \mapsto q_i, \ldots \}, q)
\]
$\SPP$s have two base cases, $\top$ and $\bot$, which represent the corresponding \NetKAT expressions. The third case of an $\SPP$ represents testing a field $f$ of an input packet against a series of values $v_0,\dots,v_i,\dots, v_n$, with a default case $f \testNE v_0 \cdots f \testNE v_n$. However, instead of continuing recursively after the test, $\SPP$s nondeterministically assign a value $w_{ij}$ to the field $f$ of the output packet, and continue recursively with the corresponding child $q_{ij}$. This way, $\SPP$s can output more than one packet for a given input packet, and can also output packets with different values for the same field. Semantically this third case of an $\SPP$ has the semantics of the \NetKAT expression:
$ \sum_i (f \test v_i \cdot \sum_j (f \mut w_{ij} \cdot q_{ij}))\ +\ \
        \left(\prod_i f \testNE v_i\right) \cdot \left(\sum_i f \mut w_i\ +\ \left(\prod_i f \testNE w_i\right) \cdot q\right)$,
where $f$ is a field, $v_i,w_{ij},w_i$ are values, and $q_{ij}, q_j, q$ are SPPs.

\begin{example}On the left, we draw the SPP corresponding to the \NetKAT expression $f\test 1 + f\mut 2$, and on the right the one for $(f \mut 0 \cdot g \mut 0) + (f\test 1 \cdot g\test 2 \cdot f\mut 3 \cdot g\mut 4)$:
\begin{center}
\input{include/spp-ex.tex}
\end{center}
\end{example}

In this paper, we write the semantics of SPPs as a function $\Sem{\cdot}: \SPP \to \Pk\times\Pk \to \bin$. This is consistent with the semantics given above as the fragment of $\Pk\cdot(\Pk\cdot\dup)^\star\cdot\Pk$ that is reachable without $\dup$ is precisely $\Pk\cdot\Pk$. We represent SPPs as directed acyclic graphs, similarly to BDDs. Vertices are labeled with the test field. Solid arrows are labeled with a number encoding the test value, and dashed arrows represent a default case. The sinks of the graph are labeled with $\top$ or $\bot$.

By removing extraneous branches and standardizing between some equivalent forms, SPPs can be made canonical in the sense that for any subset of $\Pk\cdot\Pk$ there is a unique normal-form SPP. In addition, SPPs are closed under all of the \NetKAT operations (in \Cref{fig:synsem}), as well as difference, symmetric difference, and intersection.

\subsection{The SPP Teacher}
We first develop a MAT-based framework (as in \Cref{sec:learningcanonical}) for learning SPPs. The SPP teacher holds a set of guarded strings represented by a $\dup$-free \NetKAT expression (i.e., guarded strings of length two, or, pairs of packets). As before, the teacher answers membership and equivalence queries:
\[ \memSPP: \Pk\times\Pk \to 2\qquad \equivSPP : \SPP \to 1 + \Pk\times\Pk \]
For a membership query, given a pair of packets, the teacher returns $\top$ if the pair is in its language, and $\bot$ otherwise. For an equivalence query, given a SPP $h$ the teacher returns $\top$ if $h$ has the semantics of the target language, otherwise, the teacher returns a pair of packets in the symmetric difference between $\Sem{h}$ and the target, i.e., a counterexample to the correctness of $h$.

If the teacher has an SPP for the target program, the implementation of both queries is straightforward: for a membership query we check that the given packet pair is in the semantics of the target (i.e., is a path from root to $\top$), and for the equivalence query, we take the symmetric difference between the target and the hypothesis. If the resulting SPP is $\bot$ we are done; otherwise we choose a counterexample packet pair by choosing a path to $\one$ in the resulting SPP.

\subsection{Evidence Packet Programs}

The learner we present below proceeds in a similar fashion to $\lstar$ \cite{angluin1987learning} or its variants (e.g., KV \cite{kearnsvazirani1994}). The learner maintains a data structure with all of the evidence received from the teacher, and it build successive queries and conjectures guided by this data structure. For learning SPPs, the data structure is called an Evidence Packet Program (EPP):
\begin{align*}
e \in \EPP \Coloneqq \top \mid \bot \mid F \times (V \Mapsto (V\Mapsto \EPP))
\end{align*}
\begin{wrapfigure}[12]{R}{0.18\textwidth}
\centering
	\begin{tikzpicture}[>=latex',line join=bevel,scale=0.5]
  \small
  \pgfsetlinewidth{1bp}
\begin{scope}
  \pgfsetstrokecolor{black}
  \definecolor{strokecol}{rgb}{1.0,1.0,1.0};
  \pgfsetstrokecolor{strokecol}
  \definecolor{fillcol}{rgb}{1.0,1.0,1.0};
  \pgfsetfillcolor{fillcol}
  \filldraw (0.0bp,0.0bp) -- (0.0bp,278.4bp) -- (126.8bp,278.4bp) -- (126.8bp,0.0bp) -- cycle;
\end{scope}
\begin{scope}
  \pgfsetstrokecolor{black}
  \definecolor{strokecol}{rgb}{1.0,1.0,1.0};
  \pgfsetstrokecolor{strokecol}
  \definecolor{fillcol}{rgb}{1.0,1.0,1.0};
  \pgfsetfillcolor{fillcol}
  \filldraw (0.0bp,0.0bp) -- (0.0bp,278.4bp) -- (126.8bp,278.4bp) -- (126.8bp,0.0bp) -- cycle;
\end{scope}
\begin{scope}
  \pgfsetstrokecolor{black}
  \definecolor{strokecol}{rgb}{1.0,1.0,1.0};
  \pgfsetstrokecolor{strokecol}
  \definecolor{fillcol}{rgb}{1.0,1.0,1.0};
  \pgfsetfillcolor{fillcol}
  \filldraw (0.0bp,0.0bp) -- (0.0bp,278.4bp) -- (126.8bp,278.4bp) -- (126.8bp,0.0bp) -- cycle;
\end{scope}
\begin{scope}
  \pgfsetstrokecolor{black}
  \definecolor{strokecol}{rgb}{1.0,1.0,1.0};
  \pgfsetstrokecolor{strokecol}
  \definecolor{fillcol}{rgb}{1.0,1.0,1.0};
  \pgfsetfillcolor{fillcol}
  \filldraw (0.0bp,0.0bp) -- (0.0bp,278.4bp) -- (126.8bp,278.4bp) -- (126.8bp,0.0bp) -- cycle;
\end{scope}
\begin{scope}
  \pgfsetstrokecolor{black}
  \definecolor{strokecol}{rgb}{1.0,1.0,1.0};
  \pgfsetstrokecolor{strokecol}
  \definecolor{fillcol}{rgb}{1.0,1.0,1.0};
  \pgfsetfillcolor{fillcol}
  \filldraw (0.0bp,0.0bp) -- (0.0bp,278.4bp) -- (126.8bp,278.4bp) -- (126.8bp,0.0bp) -- cycle;
\end{scope}
\begin{scope}
  \pgfsetstrokecolor{black}
  \definecolor{strokecol}{rgb}{1.0,1.0,1.0};
  \pgfsetstrokecolor{strokecol}
  \definecolor{fillcol}{rgb}{1.0,1.0,1.0};
  \pgfsetfillcolor{fillcol}
  \filldraw (0.0bp,0.0bp) -- (0.0bp,278.4bp) -- (126.8bp,278.4bp) -- (126.8bp,0.0bp) -- cycle;
\end{scope}
\begin{scope}
  \pgfsetstrokecolor{black}
  \definecolor{strokecol}{rgb}{1.0,1.0,1.0};
  \pgfsetstrokecolor{strokecol}
  \definecolor{fillcol}{rgb}{1.0,1.0,1.0};
  \pgfsetfillcolor{fillcol}
  \filldraw (0.0bp,0.0bp) -- (0.0bp,278.4bp) -- (126.8bp,278.4bp) -- (126.8bp,0.0bp) -- cycle;
\end{scope}
  \pgfsetcolor{black}
  \draw [->] (86.113bp,258.23bp) .. controls (82.799bp,252.91bp) and (78.876bp,245.73bp)  .. (76.8bp,238.8bp) .. controls (74.527bp,231.21bp) and (73.802bp,222.28bp)  .. (73.618bp,208.96bp);
  \definecolor{strokecol}{rgb}{0.0,0.0,0.0};
  \pgfsetstrokecolor{strokecol}
  \draw (82.8bp,232.8bp) node { 1};
  \draw [->] (94.925bp,256.88bp) .. controls (98.673bp,245.22bp) and (104.92bp,225.8bp)  .. (110.64bp,208.01bp);
  \draw (110.18bp,232.8bp) node { 2};
  \draw [->] (73.8bp,197.52bp) .. controls (73.8bp,188.76bp) and (73.8bp,170.0bp)  .. (73.8bp,150.08bp);
  \draw (79.8bp,174.0bp) node { 1};
  \draw [->] (71.69bp,199.53bp) .. controls (68.592bp,195.29bp) and (62.653bp,187.1bp)  .. (57.8bp,180.0bp) .. controls (52.011bp,171.53bp) and (45.714bp,161.9bp)  .. (37.277bp,148.81bp);
  \draw (63.8bp,174.0bp) node { 3};
  \draw [->] (73.8bp,127.93bp) .. controls (73.8bp,116.73bp) and (73.8bp,98.855bp)  .. (73.8bp,80.991bp);
  \draw (79.8bp,104.4bp) node { 2};
  \draw [->] (74.303bp,69.527bp) .. controls (75.262bp,61.005bp) and (77.395bp,42.056bp)  .. (79.669bp,21.85bp);
  \draw (83.51bp,45.6bp) node { 3};
  \draw [->] (30.354bp,128.21bp) .. controls (28.708bp,116.83bp) and (26.03bp,98.323bp)  .. (23.441bp,80.43bp);
  \draw (33.57bp,104.4bp) node { 2};
  \draw [->] (22.013bp,69.919bp) .. controls (20.403bp,61.575bp) and (16.658bp,42.163bp)  .. (12.771bp,22.019bp);
  \draw (24.228bp,45.6bp) node { 3};
  \draw [->] (112.02bp,197.93bp) .. controls (112.43bp,189.4bp) and (113.34bp,170.46bp)  .. (114.32bp,150.25bp);
  \draw (119.39bp,174.0bp) node { 1};
  \draw [->] (114.8bp,127.93bp) .. controls (114.8bp,116.73bp) and (114.8bp,98.855bp)  .. (114.8bp,80.991bp);
  \draw (120.8bp,104.4bp) node { 2};
  \draw [->] (113.16bp,71.0bp) .. controls (108.94bp,63.288bp) and (97.494bp,42.34bp)  .. (86.4bp,22.045bp);
  \draw (107.85bp,45.6bp) node { 4};
\begin{scope}
  \definecolor{strokecol}{rgb}{0.0,0.0,0.0};
  \pgfsetstrokecolor{strokecol}
  \definecolor{fillcol}{rgb}{0.97,0.78,0.54};
  \pgfsetfillcolor{fillcol}
  \filldraw (73.8bp,80.4bp) -- (68.4bp,75.0bp) -- (73.8bp,69.6bp) -- (79.2bp,75.0bp) -- cycle;
\end{scope}
\begin{scope}
  \definecolor{strokecol}{rgb}{0.0,0.0,0.0};
  \pgfsetstrokecolor{strokecol}
  \draw (91.6bp,21.6bp) -- (70.0bp,21.6bp) -- (70.0bp,0.0bp) -- (91.6bp,0.0bp) -- cycle;
  \draw (80.8bp,10.8bp) node {$\top$};
\end{scope}
\begin{scope}
  \definecolor{strokecol}{rgb}{0.0,0.0,0.0};
  \pgfsetstrokecolor{strokecol}
  \definecolor{fillcol}{rgb}{0.97,0.78,0.54};
  \pgfsetfillcolor{fillcol}
  \filldraw (22.8bp,80.4bp) -- (17.4bp,75.0bp) -- (22.8bp,69.6bp) -- (28.2bp,75.0bp) -- cycle;
\end{scope}
\begin{scope}
  \definecolor{strokecol}{rgb}{0.0,0.0,0.0};
  \pgfsetstrokecolor{strokecol}
  \draw (21.6bp,21.6bp) -- (0.0bp,21.6bp) -- (0.0bp,0.0bp) -- (21.6bp,0.0bp) -- cycle;
  \draw (10.8bp,10.8bp) node {$\bot$};
\end{scope}
\begin{scope}
  \definecolor{strokecol}{rgb}{0.0,0.0,0.0};
  \pgfsetstrokecolor{strokecol}
  \definecolor{fillcol}{rgb}{0.97,0.78,0.54};
  \pgfsetfillcolor{fillcol}
  \filldraw (114.8bp,80.4bp) -- (109.4bp,75.0bp) -- (114.8bp,69.6bp) -- (120.2bp,75.0bp) -- cycle;
\end{scope}
\begin{scope}
  \definecolor{strokecol}{rgb}{0.0,0.0,0.0};
  \pgfsetstrokecolor{strokecol}
  \definecolor{fillcol}{rgb}{0.04,0.78,0.65};
  \pgfsetfillcolor{fillcol}
  \filldraw (73.8bp,208.8bp) -- (68.4bp,203.4bp) -- (73.8bp,198.0bp) -- (79.2bp,203.4bp) -- cycle;
\end{scope}
\begin{scope}
  \definecolor{strokecol}{rgb}{0.0,0.0,0.0};
  \pgfsetstrokecolor{strokecol}
  \definecolor{fillcol}{rgb}{0.04,0.78,0.65};
  \pgfsetfillcolor{fillcol}
  \filldraw (111.8bp,208.8bp) -- (106.4bp,203.4bp) -- (111.8bp,198.0bp) -- (117.2bp,203.4bp) -- cycle;
\end{scope}
\begin{scope}
  \definecolor{strokecol}{rgb}{0.0,0.0,0.0};
  \pgfsetstrokecolor{strokecol}
  \definecolor{fillcol}{rgb}{0.97,0.78,0.54};
  \pgfsetfillcolor{fillcol}
  \filldraw [opacity=1] (73.8bp,139.2bp) ellipse (10.8bp and 10.8bp);
  \draw (73.8bp,139.2bp) node {g};
\end{scope}
\begin{scope}
  \definecolor{strokecol}{rgb}{0.0,0.0,0.0};
  \pgfsetstrokecolor{strokecol}
  \definecolor{fillcol}{rgb}{0.97,0.78,0.54};
  \pgfsetfillcolor{fillcol}
  \filldraw [opacity=1] (31.8bp,139.2bp) ellipse (10.8bp and 10.8bp);
  \draw (31.8bp,139.2bp) node {g};
\end{scope}
\begin{scope}
  \definecolor{strokecol}{rgb}{0.0,0.0,0.0};
  \pgfsetstrokecolor{strokecol}
  \definecolor{fillcol}{rgb}{0.97,0.78,0.54};
  \pgfsetfillcolor{fillcol}
  \filldraw [opacity=1] (114.8bp,139.2bp) ellipse (10.8bp and 10.8bp);
  \draw (114.8bp,139.2bp) node {g};
\end{scope}
\begin{scope}
  \definecolor{strokecol}{rgb}{0.0,0.0,0.0};
  \pgfsetstrokecolor{strokecol}
  \definecolor{fillcol}{rgb}{0.04,0.78,0.65};
  \pgfsetfillcolor{fillcol}
  \filldraw [opacity=1] (91.8bp,267.6bp) ellipse (10.8bp and 10.8bp);
  \draw (91.8bp,267.6bp) node {f};
\end{scope}
\end{tikzpicture}
\end{wrapfigure}

As with SPPs, we require the fields to appear in a fixed order as we descend the tree
structure of an EPP. Unlike SPPs, we will maintain that every field is present (i.e.  the $\top$ and $\bot$ leaves all have the same depth). The other distinction from SPPs is that every edge
label in an EPP has a concrete value; i.e., there are no default cases. The EPP
is essentially a trie containing example packet pairs, where the values are
paired by field, as in SPPs.

An example EPP containing the packet pair $(\{f \mapsto 1, g \mapsto 2\}, \{f \mapsto 3, g \mapsto 3\})$, labeled $\zero$, and the pairs $(\{f \mapsto 1, g \mapsto 2\}, \{f \mapsto 1, g \mapsto 3\})$ and  $(\{f \mapsto 2, g \mapsto 2\}, \{f \mapsto 1, g \mapsto 4\})$, labeled $\one$, is shown on the right.

We define the semantics of an EPP as $\Sem{\cdot} \colon \EPP \to \Pk\times\Pk \rightharpoonup \bin$, which, given a packet
pair, gives the label of the pair in the EPP.
\begin{align*}
  \Sem{\one}(\pk, \pkp) &\triangleq \one \qquad \Sem{\zero}(\pk, \pkp) \triangleq \zero\\
  \Sem{(f,e)}(\pk, \pkp) &\triangleq
        \begin{cases}
            \Sem{e[\pk.f][\pkp.f]}(\pk, \pkp) &\text{if } \pk.f\in\keys(e)
                                        \wedge \pkp.f\in\keys(e[\pk.f])\qquad\qquad\qquad\qquad\\
            \text{undefined} &\text{otherwise}
        \end{cases}
\end{align*}

The only pairs that we add to an EPP come from membership queries. Hence, we maintain by
construction the invariant that $\Sem{e}$ is always a restriction of the
semantics of the target SPP, $\Sem{s}$, in the sense that whenever
$\Sem{e}(\pk, \pkp) = \ell$, it is also the case that $\Sem{e}(\pk, \pkp) = \ell$.

\subsection{The SPP Learning Algorithm}

\begin{wrapfigure}[9]{r}{0.39\textwidth}
	\vspace{-12pt}
    \begin{algorithmic}
        \State $e \gets \zero$
        \State $c \gets \equivSPP(\zero)$
        \While {$c \neq \one$}
            \State $e \gets \update(e)(c)(\memSPP(c))$
            \State $h \gets \hypSPP(e)$
            \State $c \gets \equivSPP(h)$
        \EndWhile
        \State \textbf{return} $h$
    \end{algorithmic}
    \caption{Algorithm for learning SPPs.}\label{fig:learnspp}
\end{wrapfigure}

The learner executes the following steps in a loop:
(i) convert the EPP to an SPP (ii) conjecture an SPP, and (iii) update the EPP with the counterexample received from the teacher, or terminate if the conjecture was correct.  The pseudocode for the learner appears in \Cref{fig:learnspp}. In the rest of this section, we further detail the learner's subroutines.

The central function of the learning process the conversion function from EPPs to a hypothesis SPPs: $\hypSPP\colon \EPP \to \SPP$, see \Cref{fig:hypspp}. It works in bottom-up fashion, following the structure of the EPP. At each $(f_i, e)$,
it selects one of the bindings of $e$ to represent the default case in the SPP. It uses the same value for the default assignment case. Ideally, the selected binding will correspond to a value that classifies the largest fraction of the evidence in the EPP---a default case with this property  helps keep the conjectured SPP small.
The mechanism for making this choice is to choose the branch that results in the
smallest SPP after canonicalization. In the pseudocode below, the function $\select(v)$ builds the SPP that would result from $v$ being chosen as the default case, and $\hypSPP$ works by taking the minimum over $\select (v)$ for all keys $v$ for each level of the EPP. Note this process does not, in general, result in the \emph{smallest possible} SPP consistent with the EPP---it turns out it is NP-hard to find smallest SPPs in general (see \Cref{sec:spphard}). To quantify the size of an SPP, we use the following function $\mu\colon \SPP\to \mathbb{N}$:
\[
\mu(\one) = \mu(\zero) = 0 \qquad \mu(SPP(f,b,m,d)) = \#\text{keys, recursively, in $b,m,d$}
\]
That is, this cost function is simply a sum of the keys in the maps at each level of the SPP, which is a reasonable approximation of the memory footprint of an SPP in a memoized implementation. When we take the minimum of SPPs in the definition below, it is with respect to $\mu$. Note that $\sppsc$ is a ``smart constructor'' that performs canonicalization of the SPPs. We rely on canonicalization to remove redundant test branches that have the same behavior as the selected default case.

\begin{figure}[t]
\begin{align*}
  &\hypSPP\ \zero \triangleq \zero \qquad \hypSPP\ \one \triangleq \one\\
  &\hypSPP(\EPP(f, e)) \triangleq \min_{v\in\keys(e)} \select (v), \text{where:}\\
  &\quad \select(v) \triangleq \sppsc(f, b, m, d), \text{where:}\\
  &\quad\quad  b = \{v' \mapsto \{v'' \mapsto \hypSPP\ e' \mid v'' \mapsto e' \in m'\} \mid v'\mapsto m'\in e, v' \neq v\}\\
  &\quad\quad m = \{v' \mapsto \hypSPP\ e' \mid v' \mapsto e' \in e[v], v'\neq v\}\\
  &\quad\quad d = \begin{cases}
    \hypSPP\ e[v][v] & \text{if } v\in\keys(e) \text{ and }v\in\keys(e[v])\\
    \zero & \text{otherwise}
  \end{cases}
\end{align*}
\caption{Definition of $\hypSPP$, for generalizing EPPs to SPPs.}\label{fig:hypspp}
\end{figure}
The $\update \colon \EPP \to \Pk\times\Pk \to \bin \to \EPP$ operation used in in \Cref{fig:learnspp} just adds an example pair to the trie. We assume an implementation that, whenever $e' = \update\ e(\pk,\pkp)\ \ell$, then the label for $(\pk,\pkp)$ is updated in $e'$ and no other pairs are affected:
\[\Sem{e'}(\pk,\pkp)=\ell \wedge \forall (\pk',\pkp')\neq(\pk,\pkp):
\Sem{e}(\pk',\pkp') = \ell' \Rightarrow \Sem{e'}(\pk',\pkp') = \ell'.\]

\newsavebox{\eppa}
\sbox{\eppa}{%
  \begin{tikzpicture}[>=latex',line join=bevel,scale=0.5]
  \footnotesize
  \pgfsetlinewidth{1bp}
\begin{scope}
  \pgfsetstrokecolor{black}
  \definecolor{strokecol}{rgb}{1.0,1.0,1.0};
  \pgfsetstrokecolor{strokecol}
  \definecolor{fillcol}{rgb}{1.0,1.0,1.0};
  \pgfsetfillcolor{fillcol}
  \filldraw (0.0bp,0.0bp) -- (0.0bp,150.0bp) -- (22.8bp,150.0bp) -- (22.8bp,0.0bp) -- cycle;
\end{scope}
\begin{scope}
  \pgfsetstrokecolor{black}
  \definecolor{strokecol}{rgb}{1.0,1.0,1.0};
  \pgfsetstrokecolor{strokecol}
  \definecolor{fillcol}{rgb}{1.0,1.0,1.0};
  \pgfsetfillcolor{fillcol}
  \filldraw (0.0bp,0.0bp) -- (0.0bp,150.0bp) -- (22.8bp,150.0bp) -- (22.8bp,0.0bp) -- cycle;
\end{scope}
  \pgfsetcolor{black}
  \draw [->] (10.8bp,127.93bp) .. controls (10.8bp,116.73bp) and (10.8bp,98.855bp)  .. (10.8bp,80.991bp);
  \definecolor{strokecol}{rgb}{0.0,0.0,0.0};
  \pgfsetstrokecolor{strokecol}
  \draw (16.8bp,104.4bp) node { 1};
  \draw [->] (10.8bp,69.119bp) .. controls (10.8bp,60.358bp) and (10.8bp,41.601bp)  .. (10.8bp,21.682bp);
  \draw (16.8bp,45.6bp) node { 1};
\begin{scope}
  \definecolor{strokecol}{rgb}{0.0,0.0,0.0};
  \pgfsetstrokecolor{strokecol}
  \definecolor{fillcol}{rgb}{0.97,0.78,0.54};
  \pgfsetfillcolor{fillcol}
  \filldraw (10.8bp,80.4bp) -- (5.4bp,75.0bp) -- (10.8bp,69.6bp) -- (16.2bp,75.0bp) -- cycle;
\end{scope}
\begin{scope}
  \definecolor{strokecol}{rgb}{0.0,0.0,0.0};
  \pgfsetstrokecolor{strokecol}
  \draw (21.6bp,21.6bp) -- (0.0bp,21.6bp) -- (0.0bp,0.0bp) -- (21.6bp,0.0bp) -- cycle;
  \draw (10.8bp,10.8bp) node {$\top$};
\end{scope}
\begin{scope}
  \definecolor{strokecol}{rgb}{0.0,0.0,0.0};
  \pgfsetstrokecolor{strokecol}
  \definecolor{fillcol}{rgb}{0.04,0.78,0.65};
  \pgfsetfillcolor{fillcol}
  \filldraw [opacity=1] (10.8bp,139.2bp) ellipse (10.8bp and 10.8bp);
  \draw (10.8bp,139.2bp) node {f};
\end{scope}
\end{tikzpicture}
 }
 \newsavebox{\eppb}
\sbox{\eppb}{%
  \begin{tikzpicture}[>=latex',line join=bevel,scale=0.5]
  \footnotesize
  \pgfsetlinewidth{1bp}
\begin{scope}
  \pgfsetstrokecolor{black}
  \definecolor{strokecol}{rgb}{1.0,1.0,1.0};
  \pgfsetstrokecolor{strokecol}
  \definecolor{fillcol}{rgb}{1.0,1.0,1.0};
  \pgfsetfillcolor{fillcol}
  \filldraw (0.0bp,0.0bp) -- (0.0bp,150.0bp) -- (62.33bp,150.0bp) -- (62.33bp,0.0bp) -- cycle;
\end{scope}
\begin{scope}
  \pgfsetstrokecolor{black}
  \definecolor{strokecol}{rgb}{1.0,1.0,1.0};
  \pgfsetstrokecolor{strokecol}
  \definecolor{fillcol}{rgb}{1.0,1.0,1.0};
  \pgfsetfillcolor{fillcol}
  \filldraw (0.0bp,0.0bp) -- (0.0bp,150.0bp) -- (62.33bp,150.0bp) -- (62.33bp,0.0bp) -- cycle;
\end{scope}
  \pgfsetcolor{black}
  \draw [->] (25.183bp,129.81bp) .. controls (21.896bp,124.48bp) and (17.976bp,117.3bp)  .. (15.8bp,110.4bp) .. controls (13.464bp,102.99bp) and (12.468bp,94.292bp)  .. (11.796bp,81.022bp);
  \definecolor{strokecol}{rgb}{0.0,0.0,0.0};
  \pgfsetstrokecolor{strokecol}
  \draw (21.8bp,104.4bp) node { 0};
  \draw [->] (33.769bp,128.48bp) .. controls (37.295bp,116.94bp) and (43.149bp,97.774bp)  .. (48.611bp,79.891bp);
  \draw (48.561bp,104.4bp) node { 1};
  \draw [->] (11.722bp,69.119bp) .. controls (11.581bp,60.358bp) and (11.279bp,41.601bp)  .. (10.959bp,21.682bp);
  \draw (17.419bp,45.6bp) node { 0};
  \draw [->] (49.878bp,69.119bp) .. controls (50.019bp,60.358bp) and (50.321bp,41.601bp)  .. (50.641bp,21.682bp);
  \draw (56.33bp,45.6bp) node { 1};
\begin{scope}
  \definecolor{strokecol}{rgb}{0.0,0.0,0.0};
  \pgfsetstrokecolor{strokecol}
  \definecolor{fillcol}{rgb}{0.97,0.78,0.54};
  \pgfsetfillcolor{fillcol}
  \filldraw (11.8bp,80.4bp) -- (6.4bp,75.0bp) -- (11.8bp,69.6bp) -- (17.2bp,75.0bp) -- cycle;
\end{scope}
\begin{scope}
  \definecolor{strokecol}{rgb}{0.0,0.0,0.0};
  \pgfsetstrokecolor{strokecol}
  \definecolor{fillcol}{rgb}{0.97,0.78,0.54};
  \pgfsetfillcolor{fillcol}
  \filldraw (49.8bp,80.4bp) -- (44.4bp,75.0bp) -- (49.8bp,69.6bp) -- (55.2bp,75.0bp) -- cycle;
\end{scope}
\begin{scope}
  \definecolor{strokecol}{rgb}{0.0,0.0,0.0};
  \pgfsetstrokecolor{strokecol}
  \draw (21.6bp,21.6bp) -- (0.0bp,21.6bp) -- (0.0bp,0.0bp) -- (21.6bp,0.0bp) -- cycle;
  \draw (10.8bp,10.8bp) node {$\bot$};
\end{scope}
\begin{scope}
  \definecolor{strokecol}{rgb}{0.0,0.0,0.0};
  \pgfsetstrokecolor{strokecol}
  \draw (61.6bp,21.6bp) -- (40.0bp,21.6bp) -- (40.0bp,0.0bp) -- (61.6bp,0.0bp) -- cycle;
  \draw (50.8bp,10.8bp) node {$\top$};
\end{scope}
\begin{scope}
  \definecolor{strokecol}{rgb}{0.0,0.0,0.0};
  \pgfsetstrokecolor{strokecol}
  \definecolor{fillcol}{rgb}{0.04,0.78,0.65};
  \pgfsetfillcolor{fillcol}
  \filldraw [opacity=1] (30.8bp,139.2bp) ellipse (10.8bp and 10.8bp);
  \draw (30.8bp,139.2bp) node {f};
\end{scope}
\end{tikzpicture}
 }
 \newsavebox{\eppc}
\sbox{\eppc}{%
  \begin{tikzpicture}[>=latex',line join=bevel,scale=0.5]
  \footnotesize
  \pgfsetlinewidth{1bp}
\begin{scope}
  \pgfsetstrokecolor{black}
  \definecolor{strokecol}{rgb}{1.0,1.0,1.0};
  \pgfsetstrokecolor{strokecol}
  \definecolor{fillcol}{rgb}{1.0,1.0,1.0};
  \pgfsetfillcolor{fillcol}
  \filldraw (0.0bp,0.0bp) -- (0.0bp,150.0bp) -- (87.4bp,150.0bp) -- (87.4bp,0.0bp) -- cycle;
\end{scope}
\begin{scope}
  \pgfsetstrokecolor{black}
  \definecolor{strokecol}{rgb}{1.0,1.0,1.0};
  \pgfsetstrokecolor{strokecol}
  \definecolor{fillcol}{rgb}{1.0,1.0,1.0};
  \pgfsetfillcolor{fillcol}
  \filldraw (0.0bp,0.0bp) -- (0.0bp,150.0bp) -- (87.4bp,150.0bp) -- (87.4bp,0.0bp) -- cycle;
\end{scope}
  \pgfsetcolor{black}
  \draw [->] (35.316bp,129.46bp) .. controls (32.152bp,123.99bp) and (28.01bp,116.8bp)  .. (24.4bp,110.4bp) .. controls (19.474bp,101.66bp) and (13.94bp,91.618bp)  .. (7.1624bp,79.236bp);
  \definecolor{strokecol}{rgb}{0.0,0.0,0.0};
  \pgfsetstrokecolor{strokecol}
  \draw (30.4bp,104.4bp) node { 0};
  \draw [->] (45.415bp,129.29bp) .. controls (52.177bp,117.27bp) and (64.129bp,96.031bp)  .. (73.757bp,78.919bp);
  \draw (68.064bp,104.4bp) node { 1};
  \draw [->] (40.4bp,127.93bp) .. controls (40.4bp,116.73bp) and (40.4bp,98.855bp)  .. (40.4bp,80.991bp);
  \draw (46.4bp,104.4bp) node { 2};
  \draw [->] (4.9147bp,69.461bp) .. controls (4.3395bp,62.797bp) and (3.7623bp,50.026bp)  .. (6.4bp,39.6bp) .. controls (7.4263bp,35.543bp) and (9.1128bp,31.397bp)  .. (13.945bp,21.937bp);
  \draw (12.4bp,45.6bp) node { 0};
  \draw [->] (75.4bp,69.119bp) .. controls (75.4bp,60.358bp) and (75.4bp,41.601bp)  .. (75.4bp,21.682bp);
  \draw (81.4bp,45.6bp) node { 1};
  \draw [->] (39.209bp,70.295bp) .. controls (36.583bp,62.131bp) and (30.167bp,42.175bp)  .. (23.588bp,21.716bp);
  \draw (38.78bp,45.6bp) node { 2};
\begin{scope}
  \definecolor{strokecol}{rgb}{0.0,0.0,0.0};
  \pgfsetstrokecolor{strokecol}
  \definecolor{fillcol}{rgb}{0.97,0.78,0.54};
  \pgfsetfillcolor{fillcol}
  \filldraw (5.4bp,80.4bp) -- (0.0bp,75.0bp) -- (5.4bp,69.6bp) -- (10.8bp,75.0bp) -- cycle;
\end{scope}
\begin{scope}
  \definecolor{strokecol}{rgb}{0.0,0.0,0.0};
  \pgfsetstrokecolor{strokecol}
  \definecolor{fillcol}{rgb}{0.97,0.78,0.54};
  \pgfsetfillcolor{fillcol}
  \filldraw (75.4bp,80.4bp) -- (70.0bp,75.0bp) -- (75.4bp,69.6bp) -- (80.8bp,75.0bp) -- cycle;
\end{scope}
\begin{scope}
  \definecolor{strokecol}{rgb}{0.0,0.0,0.0};
  \pgfsetstrokecolor{strokecol}
  \definecolor{fillcol}{rgb}{0.97,0.78,0.54};
  \pgfsetfillcolor{fillcol}
  \filldraw (40.4bp,80.4bp) -- (35.0bp,75.0bp) -- (40.4bp,69.6bp) -- (45.8bp,75.0bp) -- cycle;
\end{scope}
\begin{scope}
  \definecolor{strokecol}{rgb}{0.0,0.0,0.0};
  \pgfsetstrokecolor{strokecol}
  \draw (31.2bp,21.6bp) -- (9.6bp,21.6bp) -- (9.6bp,0.0bp) -- (31.2bp,0.0bp) -- cycle;
  \draw (20.4bp,10.8bp) node {$\bot$};
\end{scope}
\begin{scope}
  \definecolor{strokecol}{rgb}{0.0,0.0,0.0};
  \pgfsetstrokecolor{strokecol}
  \draw (86.2bp,21.6bp) -- (64.6bp,21.6bp) -- (64.6bp,0.0bp) -- (86.2bp,0.0bp) -- cycle;
  \draw (75.4bp,10.8bp) node {$\top$};
\end{scope}
\begin{scope}
  \definecolor{strokecol}{rgb}{0.0,0.0,0.0};
  \pgfsetstrokecolor{strokecol}
  \definecolor{fillcol}{rgb}{0.04,0.78,0.65};
  \pgfsetfillcolor{fillcol}
  \filldraw [opacity=1] (40.4bp,139.2bp) ellipse (10.8bp and 10.8bp);
  \draw (40.4bp,139.2bp) node {f};
\end{scope}
\end{tikzpicture}
 }
 \newsavebox{\sppa}
\sbox{\sppa}{%
  \begin{tikzpicture}[>=latex',line join=bevel,scale=0.5]
  \footnotesize
  \pgfsetlinewidth{1bp}
\begin{scope}
  \pgfsetstrokecolor{black}
  \definecolor{strokecol}{rgb}{1.0,1.0,1.0};
  \pgfsetstrokecolor{strokecol}
  \definecolor{fillcol}{rgb}{1.0,1.0,1.0};
  \pgfsetfillcolor{fillcol}
  \filldraw (0.0bp,0.0bp) -- (0.0bp,126.0bp) -- (21.6bp,126.0bp) -- (21.6bp,0.0bp) -- cycle;
\end{scope}
\begin{scope}
  \pgfsetstrokecolor{black}
  \definecolor{strokecol}{rgb}{1.0,1.0,1.0};
  \pgfsetstrokecolor{strokecol}
  \definecolor{fillcol}{rgb}{1.0,1.0,1.0};
  \pgfsetfillcolor{fillcol}
  \filldraw (0.0bp,0.0bp) -- (0.0bp,126.0bp) -- (21.6bp,126.0bp) -- (21.6bp,0.0bp) -- cycle;
\end{scope}
  \pgfsetcolor{black}
  \draw [->,dashed] (10.8bp,104.04bp) .. controls (10.8bp,95.715bp) and (10.8bp,83.94bp)  .. (10.8bp,68.954bp);
  \draw [->,dashed] (10.8bp,57.003bp) .. controls (10.8bp,50.394bp) and (10.8bp,38.418bp)  .. (10.8bp,21.966bp);
\begin{scope}
  \definecolor{strokecol}{rgb}{0.0,0.0,0.0};
  \pgfsetstrokecolor{strokecol}
  \definecolor{fillcol}{rgb}{0.97,0.78,0.54};
  \pgfsetfillcolor{fillcol}
  \filldraw (10.8bp,68.4bp) -- (5.4bp,63.0bp) -- (10.8bp,57.6bp) -- (16.2bp,63.0bp) -- cycle;
\end{scope}
\begin{scope}
  \definecolor{strokecol}{rgb}{0.0,0.0,0.0};
  \pgfsetstrokecolor{strokecol}
  \draw (21.6bp,21.6bp) -- (0.0bp,21.6bp) -- (0.0bp,0.0bp) -- (21.6bp,0.0bp) -- cycle;
  \draw (10.8bp,10.8bp) node {$\top$};
\end{scope}
\begin{scope}
  \definecolor{strokecol}{rgb}{0.0,0.0,0.0};
  \pgfsetstrokecolor{strokecol}
  \definecolor{fillcol}{rgb}{0.04,0.78,0.65};
  \pgfsetfillcolor{fillcol}
  \filldraw [opacity=1] (10.8bp,115.2bp) ellipse (10.8bp and 10.8bp);
  \draw (10.8bp,115.2bp) node {f};
\end{scope}
\end{tikzpicture}
 }
 \newsavebox{\sppb}
\sbox{\sppb}{%
  \begin{tikzpicture}[>=latex',line join=bevel,scale=0.5]
  \footnotesize
  \pgfsetlinewidth{1bp}
\begin{scope}
  \pgfsetstrokecolor{black}
  \definecolor{strokecol}{rgb}{1.0,1.0,1.0};
  \pgfsetstrokecolor{strokecol}
  \definecolor{fillcol}{rgb}{1.0,1.0,1.0};
  \pgfsetfillcolor{fillcol}
  \filldraw (0.0bp,0.0bp) -- (0.0bp,139.0bp) -- (45.2bp,139.0bp) -- (45.2bp,0.0bp) -- cycle;
\end{scope}
\begin{scope}
  \pgfsetstrokecolor{black}
  \definecolor{strokecol}{rgb}{1.0,1.0,1.0};
  \pgfsetstrokecolor{strokecol}
  \definecolor{fillcol}{rgb}{1.0,1.0,1.0};
  \pgfsetfillcolor{fillcol}
  \filldraw (0.0bp,0.0bp) -- (0.0bp,139.0bp) -- (45.2bp,139.0bp) -- (45.2bp,0.0bp) -- cycle;
\end{scope}
  \pgfsetcolor{black}
  \draw [->] (13.863bp,117.91bp) .. controls (11.498bp,112.62bp) and (8.8108bp,105.79bp)  .. (7.4bp,99.4bp) .. controls (5.6941bp,91.678bp) and (5.2167bp,82.759bp)  .. (5.21bp,69.515bp);
  \definecolor{strokecol}{rgb}{0.0,0.0,0.0};
  \pgfsetstrokecolor{strokecol}
  \draw (13.4bp,93.4bp) node { 0};
  \draw [->,dashed] (20.971bp,117.21bp) .. controls (23.955bp,105.6bp) and (28.849bp,86.579bp)  .. (33.409bp,68.853bp);
  \draw [->,dashed] (34.4bp,57.904bp) .. controls (34.4bp,51.015bp) and (34.4bp,38.41bp)  .. (34.4bp,21.774bp);
\begin{scope}
  \definecolor{strokecol}{rgb}{0.0,0.0,0.0};
  \pgfsetstrokecolor{strokecol}
  \definecolor{fillcol}{rgb}{0.97,0.78,0.54};
  \pgfsetfillcolor{fillcol}
  \filldraw (5.4bp,69.4bp) -- (0.0bp,64.0bp) -- (5.4bp,58.6bp) -- (10.8bp,64.0bp) -- cycle;
\end{scope}
\begin{scope}
  \definecolor{strokecol}{rgb}{0.0,0.0,0.0};
  \pgfsetstrokecolor{strokecol}
  \definecolor{fillcol}{rgb}{0.97,0.78,0.54};
  \pgfsetfillcolor{fillcol}
  \filldraw (34.4bp,69.4bp) -- (29.0bp,64.0bp) -- (34.4bp,58.6bp) -- (39.8bp,64.0bp) -- cycle;
\end{scope}
\begin{scope}
  \definecolor{strokecol}{rgb}{0.0,0.0,0.0};
  \pgfsetstrokecolor{strokecol}
  \draw (45.2bp,21.6bp) -- (23.6bp,21.6bp) -- (23.6bp,0.0bp) -- (45.2bp,0.0bp) -- cycle;
  \draw (34.4bp,10.8bp) node {$\top$};
\end{scope}
\begin{scope}
  \definecolor{strokecol}{rgb}{0.0,0.0,0.0};
  \pgfsetstrokecolor{strokecol}
  \definecolor{fillcol}{rgb}{0.04,0.78,0.65};
  \pgfsetfillcolor{fillcol}
  \filldraw [opacity=1] (18.4bp,128.2bp) ellipse (10.8bp and 10.8bp);
  \draw (18.4bp,128.2bp) node {f};
\end{scope}
\end{tikzpicture}
 }
 \newsavebox{\sppc}
\sbox{\sppc}{%
  \begin{tikzpicture}[>=latex',line join=bevel,scale=0.5]
  \footnotesize
  \pgfsetlinewidth{1bp}
\begin{scope}
  \pgfsetstrokecolor{black}
  \definecolor{strokecol}{rgb}{1.0,1.0,1.0};
  \pgfsetstrokecolor{strokecol}
  \definecolor{fillcol}{rgb}{1.0,1.0,1.0};
  \pgfsetfillcolor{fillcol}
  \filldraw (0.0bp,0.0bp) -- (0.0bp,150.0bp) -- (61.6bp,150.0bp) -- (61.6bp,0.0bp) -- cycle;
\end{scope}
\begin{scope}
  \pgfsetstrokecolor{black}
  \definecolor{strokecol}{rgb}{1.0,1.0,1.0};
  \pgfsetstrokecolor{strokecol}
  \definecolor{fillcol}{rgb}{1.0,1.0,1.0};
  \pgfsetfillcolor{fillcol}
  \filldraw (0.0bp,0.0bp) -- (0.0bp,150.0bp) -- (61.6bp,150.0bp) -- (61.6bp,0.0bp) -- cycle;
\end{scope}
  \pgfsetcolor{black}
  \draw [->] (22.964bp,129.21bp) .. controls (20.352bp,123.9bp) and (17.352bp,116.95bp)  .. (15.8bp,110.4bp) .. controls (13.976bp,102.71bp) and (13.502bp,93.785bp)  .. (13.567bp,80.525bp);
  \definecolor{strokecol}{rgb}{0.0,0.0,0.0};
  \pgfsetstrokecolor{strokecol}
  \draw (21.8bp,104.4bp) node { 1};
  \draw [->,dashed] (30.925bp,128.48bp) .. controls (34.673bp,116.82bp) and (40.92bp,97.398bp)  .. (46.638bp,79.613bp);
  \draw [->] (13.584bp,69.527bp) .. controls (13.173bp,61.005bp) and (12.259bp,42.056bp)  .. (11.285bp,21.85bp);
  \draw (18.657bp,45.6bp) node { 1};
  \draw [->,dashed] (48.016bp,69.527bp) .. controls (48.427bp,61.005bp) and (49.341bp,42.056bp)  .. (50.315bp,21.85bp);
\begin{scope}
  \definecolor{strokecol}{rgb}{0.0,0.0,0.0};
  \pgfsetstrokecolor{strokecol}
  \definecolor{fillcol}{rgb}{0.97,0.78,0.54};
  \pgfsetfillcolor{fillcol}
  \filldraw (13.8bp,80.4bp) -- (8.4bp,75.0bp) -- (13.8bp,69.6bp) -- (19.2bp,75.0bp) -- cycle;
\end{scope}
\begin{scope}
  \definecolor{strokecol}{rgb}{0.0,0.0,0.0};
  \pgfsetstrokecolor{strokecol}
  \definecolor{fillcol}{rgb}{0.97,0.78,0.54};
  \pgfsetfillcolor{fillcol}
  \filldraw (47.8bp,80.4bp) -- (42.4bp,75.0bp) -- (47.8bp,69.6bp) -- (53.2bp,75.0bp) -- cycle;
\end{scope}
\begin{scope}
  \definecolor{strokecol}{rgb}{0.0,0.0,0.0};
  \pgfsetstrokecolor{strokecol}
  \draw (21.6bp,21.6bp) -- (0.0bp,21.6bp) -- (0.0bp,0.0bp) -- (21.6bp,0.0bp) -- cycle;
  \draw (10.8bp,10.8bp) node {$\top$};
\end{scope}
\begin{scope}
  \definecolor{strokecol}{rgb}{0.0,0.0,0.0};
  \pgfsetstrokecolor{strokecol}
  \draw (61.6bp,21.6bp) -- (40.0bp,21.6bp) -- (40.0bp,0.0bp) -- (61.6bp,0.0bp) -- cycle;
  \draw (50.8bp,10.8bp) node {$\bot$};
\end{scope}
\begin{scope}
  \definecolor{strokecol}{rgb}{0.0,0.0,0.0};
  \pgfsetstrokecolor{strokecol}
  \definecolor{fillcol}{rgb}{0.04,0.78,0.65};
  \pgfsetfillcolor{fillcol}
  \filldraw [opacity=1] (27.8bp,139.2bp) ellipse (10.8bp and 10.8bp);
  \draw (27.8bp,139.2bp) node {f};
\end{scope}
\end{tikzpicture}
 }

\begin{example} Consider applying the algorithm in \Cref{fig:learnspp} to the \NetKAT expression $f=1$. The SPPs wrapped in blue boxes below indicate the conjectures $h$ made in each iteration, while the EPPs drawn to the left of the SPPs indicate the set of evidence $e$ according to which the conjectures are made. The labels on the arrows indicate the counterexamples given by the teacher in each iteration. The initial state of the algorithm before the while loop (when $e = \bot$) is omitted for brevity.
\newline
\begin{tikzpicture}
 \node (eppa) [inner sep=4pt,outer sep=0pt] at (0,0)
 {\usebox\eppa};
 \node (sppa) [draw=blue!60!white,thick,rounded corners,inner sep=4pt,outer sep=0pt,minimum height=3cm] at (0.9,0)
 {\usebox\sppa};
 \node (eppb) [inner sep=4pt,outer sep=0pt]  at (5,0)
 {\usebox\eppb};
 \node (sppb) [draw=blue!60!white,thick,rounded corners,inner sep=4pt,outer sep=0pt,minimum height=3cm] at (6.3,0)
 {\usebox\sppb};
 \node (eppc) [inner sep=4pt,outer sep=0pt] at (10.8,0)
 {\usebox\eppc};
 \node (sppc) [draw=blue!60!white,thick,rounded corners,inner sep=4pt,outer sep=0pt,minimum height=3cm] at (12.6,0)
 {\usebox\sppc};
 \draw (sppa) edge[->] node[font=\footnotesize,above] {$(\{f\mapsto 0\},\{f\mapsto 0\}),\zero$} (eppb);
 \draw (sppb) edge[->] node[font=\footnotesize,above] {$(\{f\mapsto 2\},\{f\mapsto 2\}),\zero$} (eppc);
\end{tikzpicture}
\end{example}

The SPP learner in \Cref{fig:learnspp} eventually learns the correct SPP held by the teacher. The following lemma shows that SPPs constructed from EPPs are consistent
with all the packet pairs in the EPP:
\begin{rlemma}{select}
\label{lem:select}
  Let $e$ be an $\EPP$, with the number of fields $|F|\geq 1$. For any $v\in V$, and let $s = \SPP(f,b,m,d) = \select\ v$ for $e$. Then
  for all $(\pk,\pkp)$ such that $\Sem{e}(\pk,\pkp) = \one$, we have
$\Sem{s}(\pk, \pkp) =\one$, and
  for all $(\pk,\pkp)$ such that $\Sem{e}(\pk,\pkp) = \zero$, we have
$\Sem{s}(\pk, \pkp) =\zero$.
\end{rlemma}

In addition, the learning algorithm terminates ensuring soundness.
\begin{rtheorem}{sppterm}
\label{thm:spptermination}
The algorithm in \Cref{fig:learnspp} terminates with a correct SPP.
\end{rtheorem}

\subsection{Finding Small SPPs Is NP-hard}\label{sec:spphard}

Typically when developing learning algorithms, a learner usually tries to make conjectures as small as possible, as smaller conjectures usually generalize better and therefore result in more informative counterexamples. However, our $\hypSPP$ does not produce SPPs which are minimal in general. It turns out that finding small SPP consistent with a given EPP is NP-hard:

\begin{definition}
The Small Consistent SPP Problem, \textsc{SmallSPP}, is to determine, given an EPP $e$ and $k \geq 0$, whether there exists an SPP $s$ consistent with $e$ such that $\mu(s) \leq k$.
\end{definition}
\begin{theorem}
\textsc{SmallSPP} is NP-hard.
\end{theorem}
\begin{proofsketch}
By reduction from \textsc{IndependentSet}, which is to determine given an undirected graph $G=(V, E)$ and $k\geq 0$, whether there is a set $V'\subseteq V$ such that $|V'|\geq k$ and for all $v_1,v_2\in V'$, $(v_1,v_2)\notin E$.  Given a graph $(V,E)$, we build an EPP with fields $f_1,f_2$ and values $V\cup V\times V \cup \{1,2\}$ as follows. First define two small EPPs:
\[
      e_{\top}  = \EPP(f_2, \{\textsf{1} \mapsto \{\textsf{2} \mapsto \top\}, \textsf{2} \mapsto \{\textsf{1} \mapsto \bot\}\}) \qquad
      e_{\bot}  = \EPP(f_2, \{\textsf{1} \mapsto \{\textsf{2} \mapsto \bot\}, \textsf{2} \mapsto \{\textsf{1} \mapsto \top\}\}),
\]
Then we define our EPP by:
  \begin{align*}
      \textsf{to\_epp}(G) & = \EPP(f_1, \{v_1 \mapsto \{\{v_1, v_2\} \mapsto \textsf{h}(v_1, v_2)\,|\, v_2\in V, v_1 \neq v_2\} \,|\, v_1 \in V\})
  \end{align*}
  where $h(v_1, v_2)$ satisfies that, for any $v_1, v_2\in V$: $h(v_1, v_2) = h(v_2, v_1) = e_{\top}$ if $\{v_1, v_2\} \notin E$, and  $h(v_1, v_2) = e_{\top}$ and $h(v_2, v_1) = e_{\bot}$ otherwise.
It can be shown that every SPP consistent with this EPP corresponds to an independent set of $G$ (the independent set is the vertices not chosen for the test branches of the SPP). As a result, the smallest SPP corresponds to the largest independent set.
\end{proofsketch}

\section{Learning Symbolic \NetKAT Automata}\label{sec:learningsymbolic}

The learner in \Cref{sec:learningcanonical} is appealing for its simplicity, but it has a critical limitation that restricts its applicability in practice. If we require states to be separated by packet, then $\PNKA$s are necessarily much larger than necessary. In addition, the learning algorithm itself must repeatedly iterate through the entire space of packets to update the observation table.
To improve on this state of affairs, we now present an algorithm, \nklstar, for learning \emph{symbolic} \NetKAT automata, which uses the SPP learner of \Cref{sec:learningspp} as a subroutine. The advantage to symbolic automata is that they are more compact, as they capture behaviors across sets of packets concisely.

\subsection{Background}

SPPs encode transformations on packets in a symbolic manner. Hence, we can construct automata that use SPPs to represent the transition and observation functions for each state.

\begin{definition}\label{def:snka}
A \emph{symbolic} \NetKAT automaton ($\SNKA$) is a four-tuple $(S, s_0, \delta, \varepsilon)$ where $S$ is a finite set of states, $s_0 \in S$ is the start state, $\delta$ and $\varepsilon$ are transition and observation functions, respectively:
$$\delta\colon S \to S \to \SPP\qquad\qquad\varepsilon\colon S \to \SPP$$
There is one additional restriction to ensure that automata are deterministic: for all states
$s,s_1,s_2\in S$ such that $s_1\neq s_2$, we must have $\delta (s) (s_1) \nf{\cap} \delta (s) (s_2) = \zero$,
where $\nf{\cap}$ is intersection on SPPs.
\end{definition}
Semantically, an \SNKA $\mathcal{M} = (S, s_0, \delta, \varepsilon)$ accepts a language $\mathcal{L}(\mathcal{M}) \subseteq \GS$ containing all strings $w \in \GS$ for which $\mathcal{M}(s_0, w) = \top$, where:
	\begin{align*}
	\mathcal{M}(s, \pk\cdot\pkp) &\triangleq \Sem{\varepsilon(s)}(\pk, \pkp)\\
	\mathcal{M}(s, \pk\cdot\pkp\cdot\dup\cdot w') &\triangleq
    \begin{cases}
    \mathcal{M}(s', \pkp\cdot w') & \exists s'\in S\text{ such that } \Sem{\delta(s)(s')}(\pk, \pkp) = \one\\
    \zero & \text{otherwise}
    \end{cases}
\end{align*}

\subsection{The \SNKA Teacher}

For the \SNKA learning problem, we assume essentially the same teacher as in \Cref{sec:learningcanonical}, except that the teacher supports equivalence queries for symbolic automata instead of \PNKA.

\begin{definition}[MAT Interface]
The MAT interface for Symbolic \NetKAT automata ($\SNKA$) is implemented as a membership and an equivalence oracle:
\[
\memSNKA: \GS \to \bin \qquad \equivSNKA : \SNKA \to 1 + \GS
\]
\end{definition}

\begin{remark}
The concept of using isolated MAT-style learning subroutines for complex transition labels was first explored by \citet{argyros2018} in the context of SFAs. However, using SFA learning for $\NetKAT$ automata would only handle the transitions in a symbolic way---i.e., the state space would blow up, as with \PNKA learning. Hence, to obtain an efficient algorithm for \SNKA learning requires additional insights and data structures, as we show below.
\end{remark}

Our general strategy for the learning algorithm is to follow the canonical learning algorithm of \Cref{fig:learncanonical}, but eliminating \emph{all} of the steps where we have to iterate over the entire packet space. To review, here are the steps where the canonical learner (\Cref{fig:learncanonical}) iterates over the packet space:
\begin{itemize}
\item We initialize a table $T_\pk$ for every packet $\pk$.
\item We initialize every $E_\pk$ to include every individual packet in $\Pk$.
\item In two places we apply $\extend (T)$: Since we may think of $T_\pk$ as already having $|\Pk|$ rows and columns (i.e., of size $|S_\pk \cup S_\pk\cdot(\Pk\cdot\dup)|$ and $|E_\pk|$, respectively), then $\extend$ requires at least $|\Pk|$ membership queries because we have just added a row or column.
\end{itemize}
Instead, we will adopt a kind of lazy approach---e.g., rather than eagerly asking $|\Pk|$ many membership queries, we will make incremental guesses and wait to be corrected by a counterexample. In particular, even if we remove all three of these steps, our conjectures will still apply to the entire packet space due to use of the EPP to SPP conversion (\Cref{sec:learningspp}). Put another way, we build an automaton with the specific evidence traces that we have seen, and then generalize the automaton from there to be well-defined on all input traces. Before specifying our learning algorithm in pseudocode, we first describe the modifications we make to the core data structures.

\subsection{Partial Observation Table}
We allow the observation table to be partial---i.e., entire packet tables as well as individual entries in packet tables may be missing. Hence, at a given point in the learning process, we will only have a packet tables for the packets we have encountered in some example. In addition, the packet table function $T_\pk$ does not have information on the extended portion ($S'_\pk$), nor is $E_\pk$ initialized with all of $\Pk$. Putting these ideas together, the type of a partial packet table is as follows:

\begin{definition}[Partial Packet Table]
For a packet $\pk \in \Pk$, a packet table is a tuple $(S_\pk, E_\pk, T_\pk)$, where:
\begin{itemize}[-]
\item $S_\pk\subseteq \{ p \mid \last(p) = \alpha, p \in \Pref \}$ is a set of \emph{access prefixes}.
\item $E_\pk \subseteq \Suf $ is a set of \emph{distinguishing suffixes}.
\item $T_\pk \colon S_\pk \times E_\pk \rightarrow \bin$ is defined such that $T_\pk(s, e) = \one$ if $s \cdot e \in \mathcal{L}$, and $T_\pk(s,e) = \zero$ otherwise.
\end{itemize}
\end{definition}

\begin{definition}[Partial Observation Table]
An observation table $P$ is a set of Partial Packet Tables, $(S_\pk, E_\pk, T_\pk)$. There may or may not be a Partial Packet Table in $P$ for each particular $\pk\in\Pk$.
\end{definition}

\paragraph{Closedness and Consistency}
Along with these changes to the table structure, we will omit the closedness check in our algorithm. The reason we can avoid this check is that, since we will not be making membership queries for an extension of every prefix by one packet, we will also not have any corresponding ``lower'' rows of our packet-tables. We will only add transitions for which we have received evidence. Note that missing transitions are perfectly acceptable in a symbolic automaton---the result is simply to drop certain traces.

We will still check a form of consistency as this is needed to ensure that the hypothesis automaton is deterministic (i.e., consistency will ensure we only allow a transition to one state for a packet pair). The modification to consistency is simple: whenever we have equal rows in a partial packet table $(S_\pk, E_\pk, T_\pk)$, we only require that their successors on, say $\pkp$, are equal \emph{if} both rows exist in $(S_\pkp, E_\pkp, T_\pkp)$ (the box highlights the difference from \Cref{def:pnkaconsistent}).

\begin{definition}[Consistent Partial Observation Table]
A partial observation table $P$ is \emph{consistent}, if for each $\pk,\pkp\in\Pk$ and for each $s,s'\in S_\pk$, then
  \framebox{\parbox{4.6cm}{if both $s\cdot\pkp\cdot\dup,s'\cdot\pkp\cdot\dup\in S_\pkp$}}, and we have  $\row_\pk(s) = \row_\pk(s')$, then we must have $\row_\pkp(s\cdot\pkp\cdot\dup) = \row_\pkp(s'\cdot\pkp\cdot\dup)$.
\end{definition}

\subsection{The \SNKA Learning Algorithm}

The rest of the \SNKA learning algorithm is similar to the one given in \Cref{fig:learncanonical}. The $\update$ function we defined before (\Cref{fig:learncanonical}) almost still works except that we add both \emph{prefixes and suffixes} of each counterexample to $P$. The updated pseudocode is shown in \Cref{fig:algs-symbolic}.

\begin{figure}[t]
    \begin{subfigure}[b]{0.49\textwidth}
	\begin{algorithmic}
           \While {$P$ not consistent}
                    \State let $e\in E_\pkp$ s.t. $\row_\pk(s) = \row_\pk(s')$, \\ \;\;\;\;\;\;\; but $T_\pkp(s\cdot\pkp\cdot\dup, e) \neq T_\pkp(s'\cdot\pkp\cdot\dup, e)$.
                    \State $E_\pk \gets E_\pk \cup \{\pkp\cdot\dup\cdot e\}$
           \EndWhile
           \State \textbf{return} $\extend(P)$
	\end{algorithmic}
	\caption{Algorithm for $\textsf{mkConsistent} \colon T \rightarrow T$.}		\vspace{10pt}
\end{subfigure}
    \begin{subfigure}[b]{0.5\textwidth}
	    \centering
    \begin{algorithmic}
		\State $S \gets \{s_0\}$
		\State $\delta(s_0)(s_0) \gets \bot$
		\State $\varepsilon(s_0) \gets \bot$
		\State \textbf{return} $\mathcal{M}_\varnothing \gets (S, s_0, \delta, \varepsilon)$
		\State
    \end{algorithmic}
  \caption{$\mathcal{M}_\varnothing$ is the $\SNKA$ of the empty language.}\vspace{10pt}
    \end{subfigure}
\begin{subfigure}{.45\textwidth}
      \centering
     \begin{algorithmic}
	   \State $\mathcal{H} \gets \mathcal{M}_\varnothing;\; P \gets \{\}$
       \While {$c \gets \equivSNKA(\mathcal{H}) \in \GS$}
       		\State $P \gets \update(P, c)$
       		\State $P \gets \textsf{mkConsistent}(P)$
           \State $\mathcal{H} \gets \hypSNKA(P)$
       \EndWhile
       \State \textbf{return} $\mathcal{H}$
     \end{algorithmic}
         \caption{Algorithm for learning symbolic automata, \nklstar.}\label{fig:learnsymbolic}
\end{subfigure}
 \begin{subfigure}[b]{0.49\textwidth}
	    \centering
    \begin{algorithmic}
           \State \ 
           \State \textbf{Input: } Partial Observation table $P$, counterexample string $c\in\GS$
       \For {$s\in\Pref, e\in\Suf$ s.t. $c = s \cdot e$}
       	 \State $S_{\last(s)} \gets S_{\last(s)} \cup \{s\}$
         \State $E_{\last(s)} \gets E_{\last(s)} \cup \{e\}$
       \EndFor
       \State \textbf{return} $\extend(P)$
    \end{algorithmic}
        \caption{Subroutine \textsf{update}.}\label{fig:updatesymbolic}
    \end{subfigure}
    \vspace{8pt}
    \caption{\SNKA learning algorithm and subroutines.}\label{fig:algs-symbolic}
\end{figure}

Due to our dramatic simplification of the outer algorithm, however, the construction of our hypothesis automaton, $\hypSNKA$, is more complicated.
To separate concerns and tame the complexity somewhat, we construct our automaton in two phases.  First, we use the observation table to build a data structure we call an \emph{evidence automaton} (EA), and then we convert the EA to an \SNKA.

\subsubsection{Constructing the Evidence Automaton}\label{sec:constructea}

Starting from an observation table, we first construct an EA. An EA is like an \SNKA but uses EPPs instead of SPPs.

\begin{definition}[Evidence Automaton]
A \NetKAT evidence automaton ($EA$) is a four-tuple $(Q, q_0, \delta, \varepsilon)$, with $Q$ a finite set of states, $q_0 \in S$ the initial state, $\delta \colon Q \to Q \to \EPP$ the transition function and $\varepsilon \colon Q \to \EPP$ the observation function.\end{definition}

Now we show how to construct an evidence automaton $\mathcal{E} = (Q, q_0, \delta, \varepsilon)$ from a partial observation table $P$. The first challenge is to create the set of states for the evidence automaton.  Here, we are guided by the semantics of the canonical automaton but wish to have as few states as possible.

\paragraph{Packet-specific states}
For each packet table $(S_\pk, E_\pk, T_\pk)\in P$, we start with a set of states by inspecting the packet table $T_\pk$ the same way that they are identified in \Cref{sec:learningcanonical}.

So for each $\pk$, we define the packet-specific states for each $\pk$ we have $Q_\pk \triangleq \{ \row_\pk(s) \mid s \in S_\pk \}$.  This means that for each individual elements $q, q' \in Q_\pk$ we have prefixes $s,s'\in S$ such that $q = \row_\pk(s)$ and $q' = \row_\pk(s')$.  Moreover, if $q\neq q'$, this means that there is a suffix $e\in E_\pk$ for which $\row_\pk(s)(e) \neq \row_\pk(s')(e)$, or, equivalently, $T_\pk(s,e) \neq T_\pk(s',e)$.

Thus $Q_\pk$ are just the set of distinct rows of the packet table $(S_\pk, E_\pk, T_\pk)$ that can be distinguished by suffixes from $E_\pk$. Next, we merge these states to produce the set of states of the symbolic \NKA.

\paragraph{Merged \NKA states}\label{sec:glob}
Now that we have $Q_\pk$ for each $\pk$, we build a set $Q$, along with a map for each $Q_\pk$ into $Q$, which we denote $\glob\colon Q_\pk \to Q$. There are only two restrictions that guide how we can combine packet-specific states into global states:
\begin{enumerate}
\item There is a special $q_0\in Q$ such that for all $Q_\pk$, we have $\glob(\row_\pk\ \pk) = q_0$.
\item For each $q,q'\in Q_\pk$ then $q\neq q'$ implies $\glob(q)\neq\glob(q')$.
\end{enumerate}

The first restriction says that we must map every start state into the same global state. The second restriction says that we cannot combine states from the same $Q_\pk$, after all these are already distinguished by a suffix in $E_\pk$, and this suffix prevents their combination in the symbolic automaton.

Every combination of packet states into global states that respects these two restrictions is valid. The reason, briefly, is that we may simply guard every behavior by a complete test for each different packet in the worst case (the final SPP might not do this, because it might not be needed). Indeed, different implementations may make different merges here. Different choices for $\glob$ will result in different sizes for the SPPs in the final automaton (state-minimal $\SNKA$s are not unique).

Although it would be interesting to explore heuristics that result in smaller final SPPs, we defer that question to future work and instead merely insist that the result be minimal with respect to the number of states. To achieve this, we choose $Q$ to have a size matching the largest $|Q_\pk|$ (clearly any smaller $Q$ will violate restriction 2, above). Otherwise we assume an arbitrary choice for $\glob$.

\paragraph{Adding EA Transitions}

Having created the set of states, $Q$, we need to build an EPP to label the transitions between each pair of states to build $\delta$. To do this, we start by initializing an empty EPP for each pair of states. Next we iterate over the entire observation table adding a positive example pair to some EPP for each row in each packet table. That is, for each trace $s\cdot\pk\cdot\dup\cdot\pkp\in S_\pkp$, we add a pair $(\pk,\pkp)$ with label $\one$ to the EPP $\delta(\glob(\row_\pk(s\cdot\pk)))(\glob(\row_\pkp(s\cdot\pk\cdot\dup\cdot\pkp)))$.

\paragraph{Determinizing pairs}

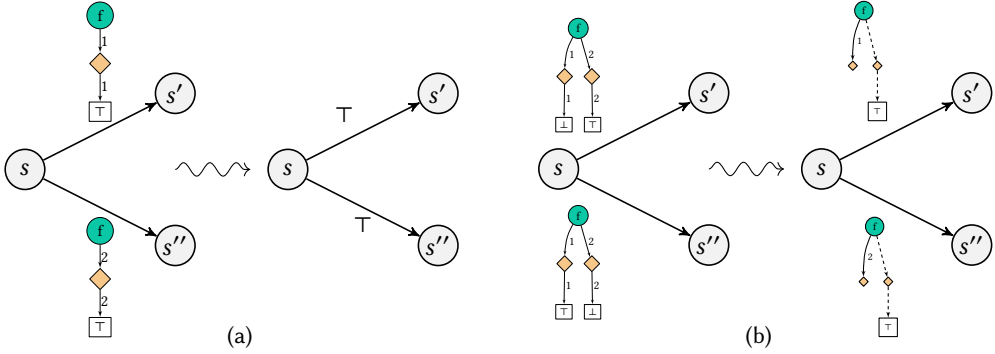
\begin{figure}[t]
	\centering
\vspace{-49pt}
    \begin{subfigure}[b]{0.49\textwidth}
	    \centering
        \begin{tikzpicture}[node distance=1.4 cm]
        \node[state] (s1) {$s$};
        \node[state] (s1') at ($(s1) + (2cm,1cm)$) {$s'$};
        \node[state] (s1'') at ($(s1) + (2cm,-1cm)$)  {$s''$};
        \newsavebox{\mybox}
        \sbox{\mybox}{%
          \begin{tikzpicture}[scale=0.45,>=latex',line join=bevel,]
  \definecolor{fillcolor}{rgb}{0.97,0.78,0.54};
  \node (n1) at (10.8bp,77.25bp) [draw,fill=fillcolor,diamond] {};
  \node (n2) at (10.8bp,10.8bp) [draw,rectangle] {$\top$};
  \definecolor{fillcolor}{rgb}{0.04,0.78,0.65};
  \node (n0) at (10.8bp,143.7bp) [draw,fill=fillcolor,circle] {f};
  \draw [->] (n0) ..controls (10.8bp,120.91bp) and (10.8bp,101.61bp)  .. (n1);
  \definecolor{strokecol}{rgb}{0.0,0.0,0.0};
  \pgfsetstrokecolor{strokecol}
  \draw (16.8bp,107.78bp) node {$1$};
  \draw [->] (n1) ..controls (10.8bp,62.114bp) and (10.8bp,42.679bp)  .. (n2);
  \draw (16.8bp,46.725bp) node {$1$};
\end{tikzpicture}
         }
        \draw (s1) edge node[above]{\scalebox{0.6}{\usebox{\mybox}}} (s1');
        \newsavebox{\thisbox}
        \sbox{\thisbox}{%
          \begin{tikzpicture}[scale=0.45,>=latex',line join=bevel,]
  \definecolor{fillcolor}{rgb}{0.97,0.78,0.54};
  \node (n1) at (10.8bp,77.25bp) [draw,fill=fillcolor,diamond] {};
  \node (n2) at (10.8bp,10.8bp) [draw,rectangle] {$\top$};
  \definecolor{fillcolor}{rgb}{0.04,0.78,0.65};
  \node (n0) at (10.8bp,143.7bp) [draw,fill=fillcolor,circle] {f};
  \draw [->] (n0) ..controls (10.8bp,120.91bp) and (10.8bp,101.61bp)  .. (n1);
  \definecolor{strokecol}{rgb}{0.0,0.0,0.0};
  \pgfsetstrokecolor{strokecol}
  \draw (16.8bp,107.78bp) node {$2$};
  \draw [->] (n1) ..controls (10.8bp,62.114bp) and (10.8bp,42.679bp)  .. (n2);
  \draw (16.8bp,46.725bp) node {$2$};
\end{tikzpicture}
         }
        \draw (s1) edge node[below]{\scalebox{0.6}{\usebox{\thisbox}}} (s1'');
        \draw[->,decorate,decoration=snake] (2,0) -> (3,0);
        \node[state, right of=s1, node distance=3.5 cm] (s2) {$s$};
        \node[state] (s2') at ($(s2) + (2cm,1cm)$)  {$s'$};
        \node[state] (s2'') at ($(s2) + (2cm,-1cm)$) {$s''$};
        \draw (s2) edge node{$\one$} (s2');
        \draw (s2) edge node[below]{$\one$} (s2'');
        \end{tikzpicture}
            \vspace{-15pt}
        \caption{}\label{fig:neg-pairs}
    \end{subfigure}
    \begin{subfigure}[b]{0.49\textwidth}
	    \centering
        \begin{tikzpicture}[node distance=1.4 cm]
        \node[state] (s1) {$s$};
        \node[state] (s1') at ($(s1) + (2cm,1cm)$) {$s'$};
        \node[state] (s1'') at ($(s1) + (2cm,-1cm)$) {$s''$};
        \newsavebox{\boxa}
        \sbox{\boxa}{%
          \begin{tikzpicture}[>=latex',line join=bevel,scale=0.6]
\begin{scope}
  \pgfsetstrokecolor{black}
  \definecolor{strokecol}{rgb}{1.0,1.0,1.0};
  \pgfsetstrokecolor{strokecol}
  \definecolor{fillcol}{rgb}{1.0,1.0,1.0};
  \pgfsetfillcolor{fillcol}
  \filldraw (0.0bp,0.0bp) -- (0.0bp,154.5bp) -- (62.36bp,154.5bp) -- (62.36bp,0.0bp) -- cycle;
\end{scope}
\begin{scope}
  \pgfsetstrokecolor{black}
  \definecolor{strokecol}{rgb}{1.0,1.0,1.0};
  \pgfsetstrokecolor{strokecol}
  \definecolor{fillcol}{rgb}{1.0,1.0,1.0};
  \pgfsetfillcolor{fillcol}
  \filldraw (0.0bp,0.0bp) -- (0.0bp,154.5bp) -- (62.36bp,154.5bp) -- (62.36bp,0.0bp) -- cycle;
\end{scope}
  \definecolor{fillcolor}{rgb}{0.97,0.78,0.54};
  \node (n1) at (11.8bp,77.25bp) [draw,fill=fillcolor,diamond] {};
  \definecolor{fillcolor}{rgb}{0.97,0.78,0.54};
  \node (n3) at (49.8bp,77.25bp) [draw,fill=fillcolor,diamond] {};
  \node (n2) at (10.8bp,10.8bp) [draw,rectangle] {$\bot$};
  \node (n4) at (50.8bp,10.8bp) [draw,rectangle] {$\top$};
  \definecolor{fillcolor}{rgb}{0.04,0.78,0.65};
  \node (n0) at (30.8bp,143.7bp) [draw,fill=fillcolor,circle] {f};
  \draw [->] (n0) ..controls (21.872bp,128.99bp) and (17.951bp,121.81bp)  .. (15.8bp,114.9bp) .. controls (13.232bp,106.65bp) and (12.26bp,96.909bp)  .. (n1);
  \definecolor{strokecol}{rgb}{0.0,0.0,0.0};
  \pgfsetstrokecolor{strokecol}
  \draw (21.8bp,107.78bp) node { 1};
  \draw [->] (n0) ..controls (37.261bp,120.78bp) and (43.318bp,100.24bp)  .. (n3);
  \draw (49.112bp,107.78bp) node { 2};
  \draw [->] (n1) ..controls (11.588bp,62.604bp) and (11.274bp,42.344bp)  .. (n2);
  \draw (17.448bp,46.725bp) node { 1};
  \draw [->] (n3) ..controls (50.012bp,62.604bp) and (50.326bp,42.344bp)  .. (n4);
  \draw (56.36bp,46.725bp) node { 2};
\end{tikzpicture}
         }
        \draw (s1) edge node[above,pos=0.3,xshift=-12pt]{\scalebox{0.45}{\usebox{\boxa}}} (s1');
        \newsavebox{\boxb}
        \sbox{\boxb}{%
          \begin{tikzpicture}[>=latex',line join=bevel,scale=0.6]
\begin{scope}
  \pgfsetstrokecolor{black}
  \definecolor{strokecol}{rgb}{1.0,1.0,1.0};
  \pgfsetstrokecolor{strokecol}
  \definecolor{fillcol}{rgb}{1.0,1.0,1.0};
  \pgfsetfillcolor{fillcol}
  \filldraw (0.0bp,0.0bp) -- (0.0bp,154.5bp) -- (62.36bp,154.5bp) -- (62.36bp,0.0bp) -- cycle;
\end{scope}
\begin{scope}
  \pgfsetstrokecolor{black}
  \definecolor{strokecol}{rgb}{1.0,1.0,1.0};
  \pgfsetstrokecolor{strokecol}
  \definecolor{fillcol}{rgb}{1.0,1.0,1.0};
  \pgfsetfillcolor{fillcol}
  \filldraw (0.0bp,0.0bp) -- (0.0bp,154.5bp) -- (62.36bp,154.5bp) -- (62.36bp,0.0bp) -- cycle;
\end{scope}
  \definecolor{fillcolor}{rgb}{0.97,0.78,0.54};
  \node (n1) at (11.8bp,77.25bp) [draw,fill=fillcolor,diamond] {};
  \definecolor{fillcolor}{rgb}{0.97,0.78,0.54};
  \node (n3) at (49.8bp,77.25bp) [draw,fill=fillcolor,diamond] {};
  \node (n2) at (10.8bp,10.8bp) [draw,rectangle] {$\top$};
  \node (n4) at (50.8bp,10.8bp) [draw,rectangle] {$\bot$};
  \definecolor{fillcolor}{rgb}{0.04,0.78,0.65};
  \node (n0) at (30.8bp,143.7bp) [draw,fill=fillcolor,circle] {f};
  \draw [->] (n0) ..controls (21.872bp,128.99bp) and (17.951bp,121.81bp)  .. (15.8bp,114.9bp) .. controls (13.232bp,106.65bp) and (12.26bp,96.909bp)  .. (n1);
  \definecolor{strokecol}{rgb}{0.0,0.0,0.0};
  \pgfsetstrokecolor{strokecol}
  \draw (21.8bp,107.78bp) node { 1};
  \draw [->] (n0) ..controls (37.261bp,120.78bp) and (43.318bp,100.24bp)  .. (n3);
  \draw (49.112bp,107.78bp) node { 2};
  \draw [->] (n1) ..controls (11.588bp,62.604bp) and (11.274bp,42.344bp)  .. (n2);
  \draw (17.448bp,46.725bp) node { 1};
  \draw [->] (n3) ..controls (50.012bp,62.604bp) and (50.326bp,42.344bp)  .. (n4);
  \draw (56.36bp,46.725bp) node { 2};
\end{tikzpicture}
         }
        \draw (s1) edge node[below, pos=0.3, xshift=-12pt]{\scalebox{0.45}{\usebox{\boxb}}} (s1'');
        \draw[->,decorate,decoration=snake] (2,0) -> (3,0);
        \node[state, right of=s1,node distance=3.5 cm] (s2) {$s$};
        \node[state] (s2') at ($(s2) + (2cm,1cm)$){$s'$};
        \node[state] (s2'') at ($(s2) + (2cm,-1cm)$) {$s''$};
        \newsavebox{\boxc}
        \sbox{\boxc}{%
          \begin{tikzpicture}[>=latex',line join=bevel,scale=0.8]
  \pgfsetlinewidth{1bp}
\begin{scope}
  \pgfsetstrokecolor{black}
  \definecolor{strokecol}{rgb}{1.0,1.0,1.0}
  \pgfsetstrokecolor{strokecol}
  \definecolor{fillcol}{rgb}{1.0,1.0,1.0}
  \pgfsetfillcolor{fillcol}
  \filldraw (0.0bp,0.0bp) -- (0.0bp,139.0bp) -- (45.2bp,139.0bp) -- (45.2bp,0.0bp) -- cycle;
\end{scope}
\begin{scope}
  \pgfsetstrokecolor{black}
  \definecolor{strokecol}{rgb}{1.0,1.0,1.0}
  \pgfsetstrokecolor{strokecol}
  \definecolor{fillcol}{rgb}{1.0,1.0,1.0}
  \pgfsetfillcolor{fillcol}
  \filldraw (0.0bp,0.0bp) -- (0.0bp,139.0bp) -- (45.2bp,139.0bp) -- (45.2bp,0.0bp) -- cycle;
\end{scope}
  \pgfsetcolor{black}
  \draw [->] (13.863bp,117.91bp) .. controls (11.498bp,112.62bp) and (8.8108bp,105.79bp)  .. (7.4bp,99.4bp) .. controls (5.6941bp,91.678bp) and (5.2167bp,82.759bp)  .. (5.21bp,69.515bp);
  \definecolor{strokecol}{rgb}{0.0,0.0,0.0}
  \pgfsetstrokecolor{strokecol}
  \draw (13.4bp,93.4bp) node { 1};
  \draw [->,dashed] (20.971bp,117.21bp) .. controls (23.955bp,105.6bp) and (28.849bp,86.579bp)  .. (33.409bp,68.853bp);
  \draw [->,dashed] (34.4bp,57.904bp) .. controls (34.4bp,51.015bp) and (34.4bp,38.41bp)  .. (34.4bp,21.774bp);
\begin{scope}
  \definecolor{strokecol}{rgb}{0.0,0.0,0.0}
  \pgfsetstrokecolor{strokecol}
  \definecolor{fillcol}{rgb}{0.97,0.78,0.54}
  \pgfsetfillcolor{fillcol}
  \filldraw (5.4bp,69.4bp) -- (0.0bp,64.0bp) -- (5.4bp,58.6bp) -- (10.8bp,64.0bp) -- cycle;
\end{scope}
\begin{scope}
  \definecolor{strokecol}{rgb}{0.0,0.0,0.0}
  \pgfsetstrokecolor{strokecol}
  \definecolor{fillcol}{rgb}{0.97,0.78,0.54}
  \pgfsetfillcolor{fillcol}
  \filldraw (34.4bp,69.4bp) -- (29.0bp,64.0bp) -- (34.4bp,58.6bp) -- (39.8bp,64.0bp) -- cycle;
\end{scope}
\begin{scope}
  \definecolor{strokecol}{rgb}{0.0,0.0,0.0}
  \pgfsetstrokecolor{strokecol}
  \draw (45.2bp,21.6bp) -- (23.6bp,21.6bp) -- (23.6bp,0.0bp) -- (45.2bp,0.0bp) -- cycle;
  \draw (34.4bp,10.8bp) node {$\top$};
\end{scope}
\begin{scope}
  \definecolor{strokecol}{rgb}{0.0,0.0,0.0}
  \pgfsetstrokecolor{strokecol}
  \definecolor{fillcol}{rgb}{0.04,0.78,0.65}
  \pgfsetfillcolor{fillcol}
  \filldraw [opacity=1] (18.4bp,128.2bp) ellipse (10.8bp and 10.8bp);
  \draw (18.4bp,128.2bp) node {f};
\end{scope}
\end{tikzpicture}
         }
        \draw (s2) edge node{\scalebox{0.4}{\usebox{\boxc}}} (s2');
        \newsavebox{\boxd}
        \sbox{\boxd}{%
          \begin{tikzpicture}[>=latex',line join=bevel,scale=0.8]
  \pgfsetlinewidth{1bp}
\begin{scope}
  \pgfsetstrokecolor{black}
  \definecolor{strokecol}{rgb}{1.0,1.0,1.0};
  \pgfsetstrokecolor{strokecol}
  \definecolor{fillcol}{rgb}{1.0,1.0,1.0};
  \pgfsetfillcolor{fillcol}
  \filldraw (0.0bp,0.0bp) -- (0.0bp,139.0bp) -- (45.2bp,139.0bp) -- (45.2bp,0.0bp) -- cycle;
\end{scope}
\begin{scope}
  \pgfsetstrokecolor{black}
  \definecolor{strokecol}{rgb}{1.0,1.0,1.0};
  \pgfsetstrokecolor{strokecol}
  \definecolor{fillcol}{rgb}{1.0,1.0,1.0};
  \pgfsetfillcolor{fillcol}
  \filldraw (0.0bp,0.0bp) -- (0.0bp,139.0bp) -- (45.2bp,139.0bp) -- (45.2bp,0.0bp) -- cycle;
\end{scope}
  \pgfsetcolor{black}
  \draw [->] (13.863bp,117.91bp) .. controls (11.498bp,112.62bp) and (8.8108bp,105.79bp)  .. (7.4bp,99.4bp) .. controls (5.6941bp,91.678bp) and (5.2167bp,82.759bp)  .. (5.21bp,69.515bp);
  \definecolor{strokecol}{rgb}{0.0,0.0,0.0};
  \pgfsetstrokecolor{strokecol}
  \draw (13.4bp,93.4bp) node { 2};
  \draw [->,dashed] (20.971bp,117.21bp) .. controls (23.955bp,105.6bp) and (28.849bp,86.579bp)  .. (33.409bp,68.853bp);
  \draw [->,dashed] (34.4bp,57.904bp) .. controls (34.4bp,51.015bp) and (34.4bp,38.41bp)  .. (34.4bp,21.774bp);
\begin{scope}
  \definecolor{strokecol}{rgb}{0.0,0.0,0.0};
  \pgfsetstrokecolor{strokecol}
  \definecolor{fillcol}{rgb}{0.97,0.78,0.54};
  \pgfsetfillcolor{fillcol}
  \filldraw (5.4bp,69.4bp) -- (0.0bp,64.0bp) -- (5.4bp,58.6bp) -- (10.8bp,64.0bp) -- cycle;
\end{scope}
\begin{scope}
  \definecolor{strokecol}{rgb}{0.0,0.0,0.0};
  \pgfsetstrokecolor{strokecol}
  \definecolor{fillcol}{rgb}{0.97,0.78,0.54};
  \pgfsetfillcolor{fillcol}
  \filldraw (34.4bp,69.4bp) -- (29.0bp,64.0bp) -- (34.4bp,58.6bp) -- (39.8bp,64.0bp) -- cycle;
\end{scope}
\begin{scope}
  \definecolor{strokecol}{rgb}{0.0,0.0,0.0};
  \pgfsetstrokecolor{strokecol}
  \draw (45.2bp,21.6bp) -- (23.6bp,21.6bp) -- (23.6bp,0.0bp) -- (45.2bp,0.0bp) -- cycle;
  \draw (34.4bp,10.8bp) node {$\top$};
\end{scope}
\begin{scope}
  \definecolor{strokecol}{rgb}{0.0,0.0,0.0};
  \pgfsetstrokecolor{strokecol}
  \definecolor{fillcol}{rgb}{0.04,0.78,0.65};
  \pgfsetfillcolor{fillcol}
  \filldraw [opacity=1] (18.4bp,128.2bp) ellipse (10.8bp and 10.8bp);
  \draw (18.4bp,128.2bp) node {f};
\end{scope}
\end{tikzpicture}
         }
        \draw (s2) edge node[below, xshift=-7pt]{\scalebox{0.4}{\usebox{\boxd}}} (s2'');
        \end{tikzpicture}
            \vspace{-15pt}
        \caption{}\label{fig:spp-diffs}
    \end{subfigure}
    \vspace{8pt}
    \caption{Converting EPP transition labels to SPP transition labels.}
\end{figure}

We would like to perform EPP to SPP conversion on each transition and observation function, but there is a fly in the ointment. Because we will convert each transition to SPP independently, two SPPs leaving the same state might generalize to ``overlap'' such that the resulting automaton is nondeterministic. For example, suppose we have two EPPs labeling two transitions leaving the same state, $s$, as shown in \Cref{fig:neg-pairs}. If we perform SPP conversion to the EA on the left, we get the symbolic \NKA on the right---which is not deterministic! Leaving state $s$ on the packet pair $(\{f\mapsto 1\},\{f \mapsto 1\})$ we can go to either $s'$ or $s''$ because $\Sem{\one}(\{f\mapsto 1\},\{f\mapsto 1\}) = \one$. To prevent this problem, we add negative examples to the EA: for each positive example pair $(\pk,\pkp)$ such that $\Sem{\delta(s)(s')}(\pk,\pkp) = \one$, we add a negative pair, $(\pk,\pkp,\zero)$ to each EPP $\delta(s)(s'')$ where $s'\neq s''$. This process requires that we do not have two of the same \emph{positive example pairs} leading to different states out of the same state, but this is ensured by the consistency check.

\paragraph{Adding EA Observations}

Defining the observation function for an \SNKA is analogous to determining whether a state in a standard DFA is accepting or not. Recall that in the canonical learner, we are guaranteed to have $\Pk\subseteq E_\pk$ which allows us to define the observation function on every final packet for each row. We perform the analogous step here. More precisely, for each $q = \row_\pk (s)$:
\[\Sem{\varepsilon (\glob (q))} (\pk,\pkp) = \ell \iff T_\pk(s, \pkp)\text{ exists and } T_\pk(s, \pkp) = \ell.\]

To summarize the EA construction: for a partial observation table $P$, we have $\ev(P) \triangleq (Q,q_0,\delta,\varepsilon)$ constructed by the following pseudocode:
\begin{algorithmic}
\State Choose $Q$ with special $q_0\in Q$ such that $|Q| = \max_{\pk} |Q_\pk|$.
\State Choose $\glob$ subject to restrictions (1) and (2) in \Cref{sec:glob}.
\For {$w \in S_\pk$ s.t. $w\cdot\pkp\cdot\dup \in S_\pkp$}
    \State Update $\delta(\glob(\row_\pk\ w))(\glob(\row_\pkp(w\cdot\pkp\cdot\dup)))$ with $(\pk,\pkp), \one$.
\EndFor
\For {$w \in S_\pk, \pkp\in E_\pk$}
	\State Update $\varepsilon(\glob(\row_\pk(w)))$ with $(\pk, \pkp), T_\pk(w,\pkp)$.
\EndFor
\end{algorithmic}

\subsubsection{Converting the Evidence Automaton to Symbolic \NetKAT Automaton}\label{sec:ea2snka}

Having constructed an EA from the Observation Table, we are close to having an \SNKA to conjecture. Conceptually, we would like to simply map our $\hypSPP$ function from \Cref{sec:learningspp} over the transition and observation EPPs in the EA, producing an \SNKA. There is one final determinization problem with doing this, which we show by way of an example.

\begin{example}\label{ex:determ}
Suppose we have added the negative example pairs to the EPPs, shown in \Cref{fig:spp-diffs}. Unfortunately, when we do the SPP conversion, we may still get the automaton on the right, which is still not deterministic! The SPPs for $f\testNE 1$ and $f\testNE 2$ overlap on, e.g., $(\{f\mapsto 3\},\{f\mapsto 3\})$.
\end{example}

This overlap illustrated by \Cref{ex:determ} can be resolved by taking differences arbitrarily (as SPP operations).  The reason is that any pairs remaining in the intersection of the semantics of two same-origin SPPs must \emph{both} not be from example pairs in the EPPs (because otherwise the overlap would have been prevented by negative pairs, above).  We can therefore compute final transition SPPs by subtracting out the other SPPs leaving the same state.

Thus, given an EA $\mathcal{E} = (Q,q_0,\delta,\varepsilon)$ we define $\sym(\mathcal{E})\triangleq (Q, q_0, \delta', \varepsilon')$, where \footnote{Using $\nf{-}$ for semantic difference of SPPs, and $\nf{\sum}$ for sum of SPPs}:
\[
\delta'(q)(q') \triangleq \hypSPP(q)(q') \nf{-} \nf{\sum}_{q''\neq q'} \hypSPP(q)(q'') \qquad \varepsilon'(q) \triangleq \hypSPP(\varepsilon(q))
\]
\subsection{Correctness of the \SNKA Learner}

We conclude by showing that our symbolic \SNKA learner is correct. The first lemma says that each conjecture is a well-defined \SNKA.

\begin{rlemma}{snkawelldef}
For any partial observation table $P$, we have that $\hypSNKA(P)$ is a well-defined \SNKA.
\end{rlemma}

\begin{rlemma}{ea2snka}
\label{lem:ea2snka-preserve}
Let $\mathcal{E} = (Q, q_0, \delta, \varepsilon)$ be an EA for a consistent partial observation table $P$ (i.e. $\mathcal{E} = \ev(P))$, and let $\mathcal{M} = (Q, q_0, \delta', \varepsilon) = \sym(\mathcal{E})$. For any $q,q'\in Q$ and any pair $(\pk,\pkp)\in\Pk\times\Pk$, then:
\begin{enumerate}[(a)]
\item \label{lem:preserve-trans} If $\Sem{\delta(q)(q')}(\pk,\pkp)=\one$, then $\Sem{\delta'(q)(q')}(\pk,\pkp) =\one$.
\item \label{lem:preserve-obs} If $\Sem{\varepsilon(q)}(\pk,\pkp)=\one$, then $\Sem{\varepsilon'(q)}(\pk,\pkp) =\one$.
\end{enumerate}
\end{rlemma}

Next we show that hypothesis automata are correct for certain examples in the table.

\begin{rlemma}{excorrect}
\label{lem:ex-correct}
Let $P$ be a consistent partial observation table for target language $\mathcal{L}\subseteq\GS$ and $\mathcal{M} = \hypSNKA(P) = (S, s_0, \delta, \varepsilon)$.
Let $u\in\GS$. If for every ``breaking point'' of $u = w \cdot e$ there is a packet table $(S_\pk,E_\pk,T_\pk)$ such that $w\in S_\pk$ and $e\in E_\pk$, then we have $\mathcal{M}(\glob(\row_\pk(w), \pk\cdot e) = \one \iff T_\pk(w, e) = \one$.
\end{rlemma}

\begin{rcorollary}{cexmonotone}
\label{lem:cex-monotone}
After receiving a counterexample $c$, any subsequent hypothesis $\mathcal{M}=(S,s_0,\delta,\varepsilon)$ is correct with respect to the target $\mathcal{L}$ on $c$.
\end{rcorollary}

\begin{rtheorem}{snkacorrect}
For a regular language $\mathcal{L}$, \nklstar (in \Cref{fig:learnsymbolic}) terminates with an \SNKA for $\mathcal{L}$.
\end{rtheorem}

\paragraph{Query complexity} It is possible to model \NetKAT using DFAs, where each packet is a character. Using this representation, the standard \lstar algorithm would learn a model using polynomial many queries in the size of this DFA. However, this representation of a \NetKAT program is in general much larger than the \SNKA learned by the algorithm in \Cref{sec:learningsymbolic}. Our formal development establishes termination of the \SNKA learner, but we leave a more precise analysis to future work. It is important to note that the two algorithms cannot be directly compared because the assumption of equivalence oracle is different for the two models.
As the symbolic representations used in SPPs and $\SNKA$s are designed to be compact in the common case, establishing complexity is likely to be more complicated than for \lstar. We note, however, that
the \PNKA learner (\Cref{sec:learningcanonical}) inherits polynomial query complexity in the size of the \PNKA by essentially the same arguments applicable to \lstar.

\subsection{Networking Example}\label{sec:netex}

We return to the network that we encoded in \Cref{sec:encoding}. As a reminder, the target expression is:
\begin{gather*}
\sw\test 1\cdot\pt\test 1\cdot((\pt\test 1\cdot\pt\mut 2 + \pt\test 2\cdot\pt\mut 1)\cdot \\
(\pt\test 1 + \pt\test 3 + \pt\test 2\cdot(\sw\test 1\cdot\sw\mut 2 + \sw\test 2\cdot\sw\mut 1))\cdot\dup)^\star\cdot \sw\test 2\cdot\pt\test 1
\end{gather*}
Our algorithm succeeds in learning a small automaton for this expression after 6 conjectures and 31 membership queries. For readability, we omit membership queries and observation tables. The SPP labels of transitions are shown using equivalent $\dup$-free expressions.

\begin{figure}[h]
 \begin{tikzpicture}[bend angle=30]
	\node[state, initial] (q0) {$q_0$};
	\draw (q0) edge[ematrix] ++ (0, -0.75);
	\node[draw=none, below of=q0, node distance=27pt] {$\zero$};
\end{tikzpicture}
\qquad
\begin{tikzpicture}[bend angle=30,node distance=2.5cm]
	\node[state, initial] (q0) {$q_0$};
	\draw (q0) edge[ematrix] ++ (0, -0.75);
	\node[draw=none, below of=q0, node distance=27pt] {$\zero$};

	\node[state, initial, right of=q0] (q1) {$q_1$};
	\draw (q1) edge[ematrix] ++ (0, -0.75);
	\node[draw=none, below of=q1, node distance=27pt] {$\one$};

    \draw (q0) edge node{$\sw\mut 2\cdot\pt\mut 2$} (q1);
    \draw (q1) edge[loop right] node{$\pt\mut 1$} (q1);
\end{tikzpicture}
\qquad
\begin{tikzpicture}[bend angle=30,node distance=2.5cm]
	\node[state, initial] (q0) {$q_0$};
	\draw (q0) edge[ematrix] ++ (0, -0.75);
	\node[draw=none, below of=q0, node distance=27pt] {$\zero$};

	\node[state, initial, right of=q0] (q1) {$q_1$};
	\draw (q1) edge[ematrix] ++ (0, -0.75);
	\node[draw=none, below of=q1, node distance=27pt] {$\pt\testNE 2$};

    \draw (q0) edge node{$\sw\mut 2\cdot\pt\mut 2$} (q1);
    \draw (q1) edge[loop right] node{$\pt\mut 1$} (q1);
\end{tikzpicture}
\caption{The first three conjectures.}\label{ex:2conj}
\end{figure}
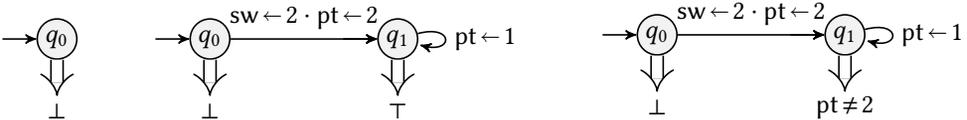

The learner starts by conjecturing the ``drop-everything'' automaton, on the right in \Cref{ex:2conj}.  The teacher responds with a positive example (valid trace):
\[
[\{\sw\mapsto 1,\pt\mapsto 1\}; \{\sw\mapsto 2,\pt\mapsto 2\}; \{\sw\mapsto 2, \pt\mapsto 1\}; \{\sw\mapsto 2,\pt\mapsto 1\}]
\]
And then the learner conjectures the automaton in the middle in \Cref{ex:2conj}.

\begin{figure}[h]
\begin{tikzpicture}[bend angle=30,node distance=2.5cm]
	\node[state, initial] (q0) {$q_0$};
	\draw (q0) edge[ematrix] ++ (0, -0.75);
	\node[draw=none, below of=q0, node distance=27pt] {$\zero$};

	\node[state, initial, right of=q0] (q1) {$q_1$};
	\draw (q1) edge[ematrix] ++ (0, -0.75);
	\node[draw=none, below of=q1, node distance=27pt] {$\pt\testNE 2$};

    \draw (q0) edge node{$\sw\mut 2\cdot\pt\mut 2$} (q1);
    \draw (q1) edge[loop right] node{$\pt\test 2\cdot\pt\mut 1$} (q1);
\end{tikzpicture}
\begin{tikzpicture}[bend angle=30,node distance=3.5cm]
	\node[state, initial] (q0) {$q_0$};
	\draw (q0) edge[ematrix] ++ (0, -0.75);
	\node[draw=none, below of=q0, node distance=27pt] {$\zero$};

	\node[state, initial, right of=q0] (q1) {$q_1$};
	\draw (q1) edge[ematrix] ++ (0, -0.75);
	\node[draw=none, below of=q1, node distance=27pt] {$\pt\testNE 2$};

    \draw (q0) edge node{$\sw\test 1\cdot \sw\mut 2\cdot\pt\mut 2$} (q1);
    \draw (q1) edge[loop right] node{$\pt\test 2\cdot\pt\mut 1$} (q1);
\end{tikzpicture}
\caption{Two further conjectures.}\label{ex:2conj-b}
\end{figure}
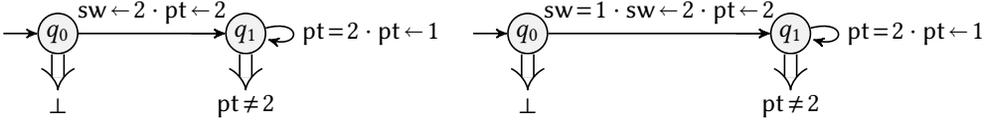

The teacher then provides a negative example  (an invalid trace): $$[\{\sw\mapsto 1,\pt\mapsto 1\};\{\sw\mapsto 2,\pt\mapsto 2\}; \{\sw\mapsto 2,\pt\mapsto 2\}]$$ and the learner conjectures the automaton on the right in \Cref{ex:2conj}.
The process continues with the teacher giving another negative example:
$$[\{\sw\mapsto 1,\pt\mapsto 1\};\{\sw\mapsto 2,\pt\mapsto 2\};
\{\sw\mapsto 2,\pt\mapsto 1\};\{\sw\mapsto 2,\pt\mapsto 1\}; \{\sw\mapsto 2,\pt\mapsto 1\}]$$
and
the learner conjecturing the automaton on the left in \Cref{ex:2conj-b}.
The teacher gives yet another negative example---
$[\{\sw\mapsto 0,\pt\mapsto 0\};\{\sw\mapsto 2,\pt\mapsto 2\}; \{\sw\mapsto 2,\pt\mapsto 1\};\{\sw\mapsto 2,\pt\mapsto 1\}]$---
followed by the learner conjecturing the automaton on the right in \Cref{ex:2conj-b}.

Finally, the teacher gives the negative example:
$$[\{\sw\mapsto 1,\pt\mapsto 0\};\{\sw\mapsto 2,\pt\mapsto 2\};\{\sw\mapsto 2,\pt\mapsto 1\};\{\sw\mapsto 2,\pt\mapsto 1\}]$$

And the learner conjectures:

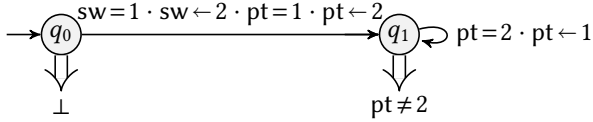
\begin{figure}[h]
\begin{tikzpicture}[bend angle=30,node distance=4.5cm]
	\node[state, initial] (q0) {$q_0$};
	\draw (q0) edge[ematrix] ++ (0, -0.75);
	\node[draw=none, below of=q0, node distance=27pt] {$\zero$};

	\node[state, initial, right of=q0] (q1) {$q_1$};
	\draw (q1) edge[ematrix] ++ (0, -0.75);
	\node[draw=none, below of=q1, node distance=27pt] {$\pt\testNE 2$};

    \draw (q0) edge node{$\sw\test 1\cdot \sw\mut 2\cdot\pt\test 1\cdot\pt\mut 2$} (q1);
    \draw (q1) edge[loop right] node{$\pt\test 2\cdot\pt\mut 1$} (q1);
\end{tikzpicture}
\caption{The sixth (and last) conjecture.}\label{ex:conj-c}
\end{figure}

This automaton is correct and the learning process terminates successfully! Though this is just a toy example, meant to illustrate the steps of the algorithm, it is worth noting that the larger networks that we deal with in the next section follow a similar structure.

\section{Implementation and Evaluation}\label{sec:evaluation}

We have prototyped our algorithms for learning SPPs (\Cref{fig:learnspp}) and $\SNKA$s (\Cref{sec:learningsymbolic}) in OCaml. Although our implementation has not yet been tuned to achieve scalable performance, it is still useful for illustrating the benefits of our approach. We have used it to experiment with learning models of real-world networks drawn from a standard benchmark.

\paragraph*{Evaluation} To evaluate our approach we sought to answer two research questions:
\begin{itemize}[-,leftmargin=*]
\item How well does the SPP learner scale with topology size for network transfer functions?
\item How does the \SNKA learner scale with topology size for whole-trace network behavior?
\end{itemize}

\paragraph{Topology Zoo Policies}
To explore these questions, we built a set of learning problem instances using the Internet Topology Zoo \cite{Knight2011}, a dataset for benchmarking network verification tools \cite{Foster2015,Moeller2024}. For each topology, we have a simple shortest-path routing policy encoded as in \Cref{sec:encoding}.

\begin{figure}[t!]
\vspace{-15pt}
  \begin{subfigure}[b]{0.3\textwidth}
    \includegraphics[width=\textwidth]{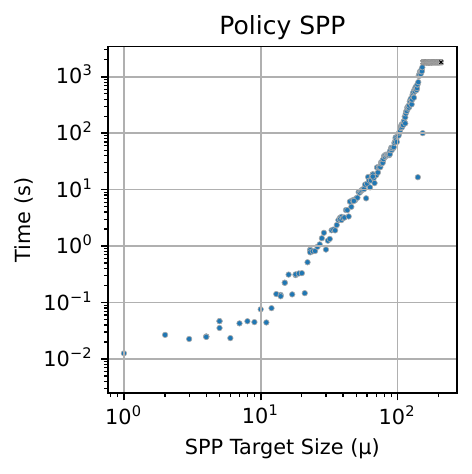}
        \vspace{-10pt}
    \caption{SPP learner timing.}
    \label{fig:spp-timing}
  \end{subfigure}
  \begin{subfigure}[b]{0.3\textwidth}
    \includegraphics[width=\textwidth]{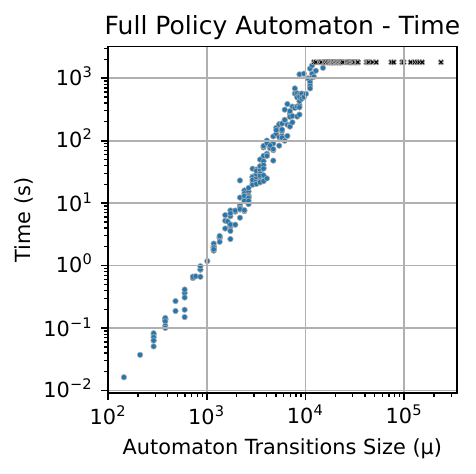}
        \vspace{-10pt}
    \caption{\SNKA learner timing.}
    \label{fig:snka-timing}
  \end{subfigure}
  \begin{subfigure}[b]{0.3\textwidth}
    \includegraphics[width=\textwidth]{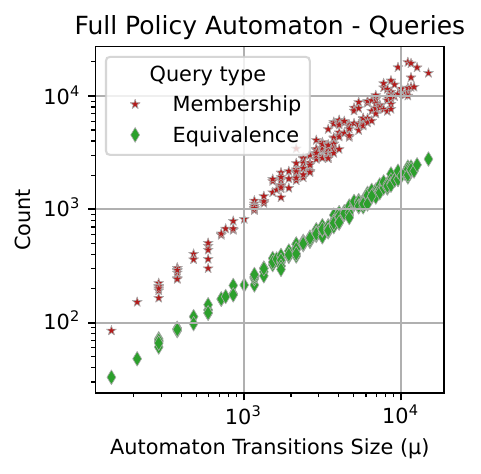}
    \vspace{-10pt}
    \caption{\SNKA learner query counts.}
    \label{fig:snka-query}
  \end{subfigure}
\vspace{8pt}
\caption{Benchmark results for Topology Zoo dataset. Examples that exceeded 30 minutes are depicted with an ``x'' in the timing plot; the queries plot includes only examples that completed within the time limit. To keep the plot legible, we omit the largest topology, KDL, as it is an outlier.}\label{fig:results}
\end{figure}

\paragraph{SPP Learning}
To evaluate the SPP learner, we set each network's ``transfer function'' ($r \cdot t$, see \Cref{sec:encoding}) as the target SPP and ran our SPP learner (that is, membership queries are on pairs of packets representing single-step transformations in the network). The results are shown in \Cref{fig:spp-timing} in log-log scales. The learner recovered the correct SPP for 190 out of 261 topologies before reaching a 30 minutes timeout, suggesting its feasibility for learning policies of modest size.

\paragraph{\SNKA Learning}
To evaluate the \SNKA learner, we instantiated the teacher with a full network behavior expression, $p_i\cdot(r\cdot t\cdot\dup)^\star\cdot p_f$.
For our input and output predicates, we choose:
$p_i \triangleq \sum_i \sw\test i \cdot (\sum_j \dst\test j)$, and
$p_f \triangleq \sum_i \sw\test i \cdot \dst\test i$.
This says that we are interested in initial packets starting at any origin with any destination, and the final packets which have reached their destination (i.e., membership queries are full-trace behaviors of the policy).  The timing results are shown in \Cref{fig:snka-timing}. The \SNKA learner succeeded on learning 190 of the 261 topologies. A plot of the number of membership queries and equivalence versus target size is shown in \Cref{fig:snka-query}. The linear shape does not mean linear behavior because the axes are log-log.

\begin{remark}
The largest topology solved by the \SNKA learner, Uunet, has 48 nodes, with 14 port numbers. Considering the fields $\sw, \dst,$ and $\pt$ used in the problem, we have $|\Pk| = 48\cdot 48\cdot 14 = 32256$. Because of the carry-on packet semantics, this means that any approach based on learning a DFA would need to identify many thousands of different \emph{states}, even if the transitions are treated symbolically, since only carry-on packets that will be dropped can be combined. Instead, the \SNKA learner here identifies an automaton with only 2 states (excluding a drop state), and even though the transition SPPs are complex, it does so after \emph{only} 15,932 membership queries, and 2775 equivalence queries.
\end{remark}

\section{Practical Concerns}

In this section we sketch additional tasks that would be needed to apply our techniques in a practical setting where the oracle is a closed-box network or network component.

\paragraph{Membership Queries}
We assume a membership oracle of type $\GS\to\bin$ because it allows us to apply the core of MAT style learning to our domain. One might observe, however, that a closed-box network might more naturally be modeled by a type like $\Pk\to\GS$ (i.e., given input packets, traces are produced), and a real system may have nondeterministic behavior. To run our algorithm on a closed-box system, one would need to implement this translation between input packets and input traces. This is not a big issue: we can approximate the $\GS\to\bin$ oracle using a ``real'' oracle by sending the first packet of our desired query trace (perhaps multiple times) and then checking whether the query trace is produced by the system in any of the trials (caching the results for later queries).

\paragraph{Equivalence Queries}
The assumption of an equivalence oracle in work based on Angluin's \lstar algorithm, including ours, is indeed a strong assumption. However, in practice, it often suffices to approximate equivalence. There is a large body of prior work (e.g., CacheQuery from PLDI '20 \cite{Vila2020}), showing that approximation is effective, so the assumption of an equivalence oracle is less of a practical hurdle than it might seem.

In many systems, equivalence queries are approximated using membership queries. In a nutshell, the idea is to first generate a large set of traces (e.g., all traces up to a given length, or a random sample of such traces) and then execute them on the conjectured model and the system. If no discrepancies are found, then the equivalence query is deemed to be successful. On the other hand, if a discrepancy is found, then the equivalence query is deemed to fail, and the trace that triggered the discrepancy is returned as a counter-example. The Random Wp-Method \cite{Fujiwara1991} and domain-specific approaches \cite{Kruger2024} offer reasonable approximation guarantees.

\section{Related Work}

This paper is the first to address symbolic learning of \NetKAT automata, however, a number of similar problems have been the topic of extensive study.

\noindent{\textbf{Learning for GKAT.}}
\citet{Zetzsche2022} adapt Angluin's \lstar to the setting of Guarded Kleene Algebra with Tests. Their work bears some similarity to ours in that they also establish a Myhill-Nerode Theorem for GKAT en route to the full presentation of their algorithm. Unfortunately, we cannot apply their results beyond inspiration because of the differences between GKAT and \NetKAT: GKAT supports only a ``guarded'' choice operator, which is deterministic, while \NetKAT supports the full nondeterministic choice operation of KAT. Further, \NetKAT's carry-on packet semantics is the source of the challenges that we solve in our paper, issues that have no analog in GKAT.  The incompatibility of \NetKAT and GKAT is explored in Wasserstein's Master's thesis~\cite{wasserstein_guarded_2023}.

\noindent\textbf{Learning OBDDs.}
Angluin-style learning of Ordered Binary Decision Diagrams (OBDDs)~\cite{Gavalda1995} and of automata with decision diagram transitions~\cite{Maler2017} have been explored. BDDs also appear in the use of \lstar to learn Java interface specifications~\cite{Alur2005} . The structure we use, Symbolic Packet Programs, are reminiscent of decision diagrams, yet have crucial differences which are essential to encode \NetKAT policies as discussed in~\cite{Moeller2024}, which require the design of a new learning algorithm.

\noindent\textbf{Learning Networking Protocols.}
 \citet{Brostean2014} successfully applied automata learning to learn the state machine of fragments of the TCP protocol, and identified bugs in a widely used TCP implementation. This work was further extended to learning sliding window behavior \cite{Brostean2017Sliding}, learning SSH implementations \cite{Brostean2017SSH},  and DTLS implementations \cite{Brostean2020}. \citet{Ferreira2021} introduced the Prognosis tool to apply automata learning to the QUIC protocol, a promising new network protocol aiming to fully replace TCP, TLS/SSL, and HTTP for the modern web. Prognosis was used to identify issues with both the QUIC RFC itself and several implementations. These approaches are prime examples of using automata learning to identify bugs in networking, but they have mostly explored DFAs, which have the drawbacks mentioned above. Our paper brings $\NetKAT$ automata as a more expressive learning target to the learning toolkit for networking.

\section{Conclusion}
In this paper we presented novel techniques for active learning of $\NetKAT$ models. We developed the theory of \NetKAT with a characterization of canonical automata (akin to the results of Myhill-Nerode for classical automata) and then used this notion to develop a learning algorithm. We designed two symbolic learning algorithms, one for the $\dup$-free fragment of \NetKAT (SPP learner) and one for the general case of symbolic $\NetKAT$ automata (\SNKA learner). We implemented these algorithms in an OCaml prototype and performed an evaluation using existing benchmarks. Our evaluation shows that both the SPP and the \SNKA learners scale with topology size.

As a natural next step, we would like to apply our algorithm to practical scenarios, including inferring the model of a device for which no configuration or program is known and analyze potential misconfigurations or malicious behavior. We would also like to explore further improvements to the algorithm via \emph{symbolic packet tables}. Currently, if multiple packets are observed to behave equivalently, we still treat their packet tables independently. This leads to repetition in queries, and a higher memory footprint than potentially needed. We would also like to investigate efficient heuristics for implementations of the state grouping function $\glob$. Currently the properties of this function focus on folding the state space, however transition SPP minimality would be as important for the scalability of $\NetKAT$ procedures that use the models.

\section*{Acknowledgements}

We are grateful to our PLDI reviewers and shepherd who helped us improve our paper significantly.
We also thank David Darais, Caleb Lucas-Foley, and Cole Schlesinger of Galois Inc. for helpful conversations on our early progress and to the Cornell PLDG for their feedback. We are also grateful to the EPFL DCSL group for providing a welcoming and supportive environment.
This work was supported in part by ONR grant N68335-22-C-0411, DARPA grant W912CG-23-C-0032, ERC grant Autoprobe (no. 101002697), and a Royal Society Wolfson fellowship, as well as gifts from Google, InfoSys, and the VMware University Research Fund.

\section*{Data Availability}

A snapshot of the OCaml code which underwent artifact evaluation is available as a dependencies-included Docker image on Zenodo \cite{Moeller2025Artifact}.

\bibliographystyle{ACM-Reference-Format}
\bibliography{refs}

\ifthenelse{\boolean{isExtendedVersion}}{%
\clearpage
\appendix
\section{Omitted Proofs}

\subsection{$\PNKA$ Properties \& Correctness}

\repeattheorem{nkaiffpnka}
\begin{proof}
\item[]($\Rightarrow$)
Let $\mathcal{N} = (S, s_0, \delta, \varepsilon)$ be an \NKA accepting $\mathcal{L}$.
We define a \PNKA $\mathcal{P} = (Q, q_0, \partial, \lambda)$, where:
$$Q = \Pk \times S\qquad
q_0(\pk) = (\pk, s_0)\qquad
\partial_{\pkp} (\pk, s) = (\pkp, \delta_{\pk\pkp}(s))\qquad
\lambda_{\pkp} (\pk, s) = \varepsilon_{\pk \pkp} (s)$$
We need to show $\mathcal{L}(\mathcal{P}) = \mathcal{L}(\mathcal{N})$. We show a stronger result: for any $s \in S$, it holds that $\mathcal{P}((\pk,s), w) \iff \mathcal{N}(s,\pk \cdot w)$. The result then follows by taking $s = s_0$. We prove the above result by induction on the length of $w$:
if $w=\pkp$, then we have $\mathcal{P}((\pk,s), \pkp) \iff \lambda_{\pkp}(\pk, s) \iff \varepsilon_{\pk\pkp}(s) \iff \mathcal{N}(s,\pk \cdot \pkp)$. On the other hand if
$w = \pkp\cdot\dup\cdot w'$, then:
\begin{align*}
\mathcal{P}((\pk,s), w) &\iff \mathcal{P}(\partial_\pkp(\pk, s), w') \iff \mathcal{P}((\pkp, \delta_{\pk\pkp}(s)), w') \\
&\stackrel{\text{IH}}\iff \mathcal{N}(\delta_{\pk\pkp}(s), \pkp\cdot w') \iff \mathcal{N}(s, \pk\cdot w).
\end{align*}

\item[]($\Leftarrow$)
Let $\mathcal{P} = (Q, q_0, \partial, \lambda)$ be a \PNKA accepting $\mathcal{L}$.
We define an \NKA $\mathcal{N} = (S, s_0, \delta, \varepsilon)$, where:
\begin{itemize}[-]
\item $S = Q \uplus \{ s_0, s_\bot \}$
\item $\delta_{\pk\pkp} (s) =
            \begin{cases}
            \partial_{\pkp} (q_0(\pk))& \text{if }s = s_0 \\
            \partial_{\pkp}(s) &  \spell(s) = \pk\\
            s_\bot & \text{otherwise}
            \end{cases}$
\qquad$\varepsilon_{\pk\pkp} (s) =
            \begin{cases}
            \lambda_{\pkp}(q_0(\pk))& \text{if }s = s_0\\
             \lambda_{\pkp}(s) &  \spell(s) = \pk\\
            \bot & \text{otherwise}
            \end{cases}$
\end{itemize}

Again, we need to show that $\mathcal{L}(\mathcal{N}) = \mathcal{L}(\mathcal{P})$. That is: $\mathcal{N}(s_0, \pk\cdot w) \iff \mathcal{P}(q_0(\alpha), w)$. We first observe that:
\[
\mathcal{N}(s_0, \pk\cdot \pkp) \iff \varepsilon_{\pk\pkp} (s_0) \iff  \lambda_{\pkp}(q_0(\pk)) \iff \mathcal{P}(q_0(\pk), \pkp).
\]
and
\begin{align*}
\mathcal{N}(s_0, \pk\cdot \pkp \cdot \dup \cdot w') &\iff  \mathcal{N}(\delta_{\pk\pkp}(s_0),  \pkp \cdot w') \iff  \mathcal{N}(\partial_{\pkp}(q_0(\alpha)),  \pkp \cdot w') \\
&\stackrel \dagger \iff
\mathcal{P}(\partial_{\pkp}(q_0(\alpha)), w') \iff \mathcal{P}(q_0(\pk),\pkp \cdot \dup  \cdot w').
\end{align*}
We conclude the proof with a proof of $\dagger$. We show a more general fact: for $q\in Q$ and $\alpha\in \Pk$, with $\spell(q)=\alpha$, it holds that $\mathcal{P}(q, w) \iff \mathcal{N}(q,  \pk \cdot w)$ (to obtain $\dagger$ instantiate this with $q= \partial_\pkp(q_0(\pk))$, $\pk = \pkp$, and $w = w'$). This fact follows by induction on the length of $w$: if $w=\pkp$, then we have $\mathcal{P}(q, \pkp) \iff \lambda_{\pkp}(q) \iff \varepsilon_{\pk\pkp}(q) \iff  \mathcal{N}(q,  \pk \cdot \pkp)$. On the other hand, let
$w = \pkp\cdot\dup\cdot w'$, and observe that $\spell(q)=\pk$ means that $\delta_{\pk\pkp}(q) =  \partial_{\pkp}(q)$. Then:
\begin{align*}
\mathcal{P}(q, w) \iff
\mathcal{P}(\partial_{\pkp}(q), w') \stackrel{\text{IH}}\iff
\mathcal{N}(\delta_{\pk\pkp}(q), \pkp\cdot w') \iff
\mathcal{N}(q, \pk\cdot w).
\end{align*}
\end{proof}

\begin{rtheorem}{quotient}
\label{thm:quotient}
	The $\PNKA$ $\mathcal{P}_{\NKmne}$ is a minimal acceptor of the language $\mathcal{L}$.
\end{rtheorem}
\begin{proof}
	For readability, let $\mathcal{P} \triangleq \mathcal{P}_{\NKmne}$. We will prove that $w = \alpha \cdot z \cdot \beta \in \GS$ is accepted by $\mathcal{P}$ if and only if $w \in \mathcal{L}$.

We first show by induction that, for a given $s \in \Pref$ we have that $\mathcal{P}([s]_{\NKmne}, z \cdot \beta) = (s \cdot z \cdot \beta \in \mathcal{L})$.

For $z$ with 0 $\dup$s, we have $\mathcal{P}([s]_{\NKmne}, \pkp) = \lambda_{\pkp}([s]_{\NKmne}) = s \cdot \beta \in \mathcal{L}$. Now let $z$ have $n$ dups, and assume that $\mathcal{P}([s]_{\NKmne}, z \cdot \pkp) = s \cdot z \cdot \pkp \in \mathcal{L}$.

Then for arbitrary $\pkpp$ we have that:
\[
\mathcal{P}([s]_{\NKmne}, \gamma \cdot \dup \cdot z \cdot \pkp) = \mathcal{P}(\partial_\gamma([s]_{\NKmne}), z \cdot \pkp) = \mathcal{P}([s \cdot \gamma \cdot \dup]_{\NKmne}, z \cdot \pkp) \stackrel{\text{IH}}{=} s \cdot \pkpp \cdot \dup \cdot z \cdot \pkp \in \mathcal{L}.
\]

We then know that $\mathcal{P}(q_0(\alpha), z \cdot \beta) = \pk \cdot z \cdot \pkp \in \mathcal{L}$, and thus $w$ is accepted by $\mathcal{P}$ iff $w \in \mathcal{L}$.

To see that $\mathcal{P}_{\NKmne}$ is minimal, observe that were $\mathcal{P}_{\NKmne}$ not minimal, that would mean there exist at least two different states $s$ and $s'$ with the same behavior that could be merged. However, were this the case, the prefixes reaching $s$ and $s'$ would then be in the same equivalence class of $\NKmne$. This contradicts the assumption that $s \neq s'$.

Therefore $\mathcal{P}_{\NKmne}$ is a minimal acceptor of $\mathcal{L}$ with $|Q| = |\{[s]_{\NKmne} \mid s \in \Pref \}| = |\NKmne|$ states.
\end{proof}

\begin{definition}[State Equivalence Relation]
	Given a \PNKA $\mathcal{M} = (Q, q_0, \partial, \lambda)$ we say that two prefixes $s$ and $t$ in $\Pref$ are indistinguishable by $\mathcal{M}$ if and only if  the state reached by the automaton on input $s$ is the same as the one reached on input $t$. Formally we define the relation:
	\[ s \equiv_{\mathcal M} t \stackrel\triangle\iff \partial^+(s) = \partial^+(t) \]

Where $\partial^+ \colon \Pref \rightarrow Q$ is defined by:
\[
\partial^+(\pk) = q_0(\pk)
\qquad
\partial^+(\pk \cdot (\pkp \cdot \dup) \cdot w) = \partial^\star(q_0(\pk), (\pkp \cdot \dup) \cdot w)
\]

And $\partial^\star \colon Q \times (\Pk \cdot \dup)^\star \rightarrow Q$ is given by:
\[
\partial^\star(q, \epsilon) = q
\qquad
\partial^\star(q, (\beta \cdot \dup) \cdot w) = \partial^\star(\partial_\pkp(q), w)
\]

Note that $\equiv_{\mathcal M}$ is an equivalence relation and that it can only have a finite number of equivalence classes, one per state.
\end{definition}

We now want to show that given a regular language $\mathcal L \subseteq \GS$ the relation $\equiv_\mathcal L$ has finite index. We do this with the help of the following lemma.

\begin{rlemma}{refinement}
\label{thm:refinement}
If $L=\mathcal L(\mathcal M)$ for a \PNKA $\mathcal M$ then for any $s, t \in \Pref$: if $s \equiv_{\mathcal M} t$ then $s \equiv_{\mathcal L} t$.
\end{rlemma}

\begin{proof}
First note that a guarded string $w \cdot \beta \in \GS$ satisfies:
\[ w \cdot \beta \in \mathcal{L}(\mathcal{M}) \iff \lambda_\beta(\partial^+(w)) = \top \]
Suppose also that $s \equiv_{\mathcal M} t$ and let $w = \alpha \cdot e \cdot \beta \in \GS$. It is easy to see that  $\partial^+(s \cdot e) =  \partial^+(t \cdot e)$. We can then reason:
\[
s \mathbin{\blacklozenge} w \in \mathcal{L} \iff \lambda_\beta(\partial^+(s \cdot e)) =\top  \iff \lambda_\beta(\partial^+(t \cdot e)) = \top \iff t \mathbin{\blacklozenge} w \in \mathcal{L}.\qedhere
\]
\end{proof}

This lemma says that if two prefixes lead to the same state in a $\PNKA$ then they are in the same equivalence class of $\NKmne$. This means that each equivalence class of $\equiv_{\mathcal L}$ is a union of equivalence classes of $\equiv_{\mathcal M}$. Therefore if $\mathcal{L}$ is regular, then $\NKmne$ has finite index. In the other direction, if $\NKmne$ has finite index, then we can build the $\PNKA$ $\mathcal{P}_{\NKmne}$ accepting $\mathcal{L}$ by \Cref{thm:quotient}, so $\mathcal{L}$ is regular.

We finish by isolating claim (3): That the minimality of $(\mathcal{P}_{\NKmne}) = (Q, q_0, \partial, \lambda)$ is \emph{unique}, and any other $\PNKA$ $\mathcal{M} = (Q', q_0', \partial', \lambda')$ with language $\mathcal{L}$ and $|\NKmne|$ states must be isomorphic to it.

\begin{rtheorem}{pnkauniqueness}
    Let $\mathcal{P} = (Q, q_0, \partial, \lambda)$ have language $\mathcal{L}$.
	If a $\PNKA$ $\mathcal{M} = (Q', q_0', \partial', \lambda')$ also has language $\mathcal{L}$ and $|\NKmne|$ states then $\mathcal{M} \cong \mathcal{P}_{\NKmne}$.
\end{rtheorem}
\begin{proof}
	For any state $q' \in Q'$ there exists a prefix $s \in \Pref$ that is an access prefix for $q'$. Formally, $\partial'^+(s) = q'$. This is guaranteed as $\mathcal{M}$'s minimality means it cannot have unreachable states.

	We now define $h \colon Q \rightarrow Q'$ by $h([s]_{\NKmne}) = \partial'^+(s)$ and show that it is a homomorphism with respect to $q_0$, $\partial$, and $\lambda$.

	The start state $q_0$ is preserved by $h$ as for any $\pk \in \Pk$, then $h(q_0(\alpha)) = \partial'^+(\pk) = q_0'(\pk)$.

	The automaton transition structure is preserved as for any $s \in \Pref$, then:
	\[h(\partial_\pkp([s]_{\NKmne})) = h([s \cdot \beta \cdot \dup]_{\NKmne}) = \partial'^+(s \cdot \beta \cdot \dup) = \partial'_\beta(\partial'^+(s)) = \partial'_\beta(h([s]_{\NKmne}))\]

	Finally, the observation function is preserved by $h$ as for any $\pkp \in \Pk$, then:
	\[\lambda_\beta([s]_{\NKmne}) \iff s \cdot \beta \in L \iff \lambda'_\pkp(\partial'^+(s)) \iff \lambda'_\beta(h([s]_{\NKmne}))\]

	We conclude the proof by showing that $h$ is actually an isomorphism. $h$ is injective as:
	\[h([s]_{\NKmne}) = h([t]_{\NKmne}) \implies \partial'^+(s) = \partial'^+(t) \implies s \equiv_{\mathcal{M}} t \implies s \equiv_{\mathcal{L}} t \implies [s]_{\NKmne} = [t]_{\NKmne}\]

	As $|Q|=|Q'|$ and $h$ is injective, $h$ is also surjective, and therefore a bijection. As such, $h$ is an isomorphism, and $\mathcal{M} \cong \mathcal{P}_{\NKmne}$.
\end{proof}

\subsection{$\PNKA$ Learning Correctness}

\repeatlemma{matchestable}
\begin{proof}
We need to justify why $q_0, \delta,$ and $\varepsilon$ are well-defined functions.
    \begin{itemize}[-]
      \item The start state selector $q_0$ is well defined because $\pk$ is added to each $S_\pk$ during initialization, so there is a $\row_\pk (\pk)$.
      \item The transition function is well defined because the rows in each $S_\pk$ are either distinct or consistent for a given transition label $\pkp$. This is guaranteed by the table's consistency. Additionally each successor (i.e., $s \cdot (\pkp\cdot\dup)$ for each $s$) is guaranteed to exist in $S'_\pkp$, and is guaranteed to have an equivalent state representative in $S_\pkp$ by the table's closedness.
      \item Each $E_\pk$ is initialized to be all of $\Pk$, so all possible $T_\pk(\cdot, \pkp)$ needed to define $\varepsilon$ are present.
    \end{itemize}
\end{proof}

\repeatlemma{hmin}
\begin{proof}
  ($\Rightarrow$) Suppose $s \equivH t$, and let $e\in\Suf$. Let $s = \pk \cdot s'$ for some $\pk$ and $s'$, and $t = \pkp \cdot t'$ for some $\pkp$ and $t'$. Clearly $\mathcal{H}(\partial^+(s), e)$ iff $\mathcal{H}(\partial^+(t), e)$. Then we have $\mathcal{H}(q_0(\pk), s \cdot e)$ iff $\mathcal{H}(q_0(\pkp), t \cdot e)$. Since $e$ is arbitrary, this means $s \equivLH t$.

($\Leftarrow$) Suppose $s \nequiv_{\mathcal{H}} t$. Then either $s$ and $t$ end with different packets (in which case clearly $s \nequiv_{L(\mathcal{H})} t$), or they both end in some packet $\pk$. As the access prefixes lead to different states, by definition of the states of $\mathcal{H}$, we must have that $(\pk, \row_\pk(s')) \neq (\pk, \row_\pk(t'))$, for some equivalent access prefixes $s'$ and $t'$ of $s$ and $t$, respectively. I.e., $s \equiv_{\mathcal{H}} s'$ and $t \equiv_{\mathcal{H}} t'$. As the rows $\row_\pk(s')$ and $\row_\pk(t')$ are different, it must be that there is a  distinguishing suffix $e\in E_\pk$ making them unequal. This suffix witnesses $s' \nequiv_{\mathcal{L}(\mathcal{H})} t'$, and therefore $s \nequiv_{\mathcal{L}(\mathcal{H})} t$.
\end{proof}

\repeatlemma{lrefine}
\begin{proof}
(By contrapositive) Suppose $s \nequivLH t$. By \Cref{lem:hmin}, this means $s \nequivH t$.
  If this is because $s$ and $t$ have different final packets, then clearly $s \nequiv_{\mathcal{L}} t$ already. Otherwise, suppose $\last(s)=\last(t)=\pk$. Then $\partial^+(s) \neq \partial^+(t)$ means $\row_\pk(s') \neq \row_\pk(t')$ for some $s', t' \in \Pref$ such that $s \equiv_{\mathcal{H}} s'$ and $t \equiv_{\mathcal{H}} t'$. This means there is an $e\in E_\pk$ for which $T_\pk(s',e) \neq T_\pk(t',e)$. The table entries come only from membership queries, and thus $e$ is a witness to $s' \nequiv_{\mathcal{L}} t'$. As $s$ and $t$ are state-equivalent to $s'$ and $t'$, respectively, by construction of the transition function of $\mathcal{H}$ then too $s \nequiv_{\mathcal{L}} t$ via the same witnessing suffix $e \in E_\pk$.
\end{proof}

\begin{rtheorem}{as}
\label{thm:as}
Given an observation table $T$, its hypothesis $\mathcal{H}$, we have that for every prefix $p \in \bigcup_{\pk\in\Pk} (S_\pk \cup S'_\pk)$ we have that $\partial^+(p) = (\last(p), \row_{\last(p)}(p))$.
\end{rtheorem}
\begin{proof}
We prove this theorem by induction on the length of $p$. Let $p$ have zero $\dup$s. Then $p = \pk$ for some $\pk\in\Pk$, and $\partial^+(\pk) = q_0(\pk) = (\pk, \row_\pk(\pk))$. Now assume that for any $p \in \bigcup_{\pk\in\Pk} (S_\pk \cup S'_\pk)$ with at most $n$ $\dup$s we have that $\partial^+(p) = (\last(p), \row_{\last(p)}(p))$.

Consider a prefix $t \in \bigcup_{\pk\in\Pk} (S_\pk \cup S'_\pk)$ with $n + 1$ $\dup$s. Then $t = p \cdot (\pk \cdot \dup)$, for some prefix $p$ and packet $\pk$. Notice now that $p$ itself is also in the set of table access prefixes, as $\bigcup_{\pk\in\Pk} (S_\pk \cup S'_\pk)$ is prefix-closed by definition. Therefore $\partial^+(t) = \partial(\partial^+(p), \pk)$. By the inductive hypothesis, this is equal to $\partial((\last(p), \row_{\last(p)}(p)), \pk) = (\pk, \row_\pk(p \cdot (\pk \cdot \dup))) = (\last(t), \row_{\last(t)}(t))$.
\end{proof}

\begin{rlemma}{tclosed}\label{lem:tclosed}
For each closed, consistent observation table $T$ produced by the PNKA learner,
and for any $\pk,\pkp\in\Pk$, if $\pkp\cdot\dup\cdot e\in E_\pk$ then $e\in E_\pkp$.
\end{rlemma}
\begin{proof}
The claim is initially trivially satisfied because all $E_\pk$ are initialized to contain dup-free suffixes.
We only add to $E_\pkp$ in $\mathsf{mkConsistent}$. By inspection, we see that we only add a trace $\pkp\cdot\dup\cdot e$ to $E_\pk$ if $e\in E_\pkp$, maintaining the invariant and proving the claim.
\end{proof}

\repeattheorem{agree}
\begin{proof}
Let $\pk\cdot s\in S_\pkpp$ and $e\in E_\pkpp$. We proceed by induction on the number of dups in $e$.

For the base case, $e = \pkp \in\Pk$. Then $\mathcal{H}(q_0(\pk), s\cdot e) = \lambda(\partial^+(\pk\cdot s),\pkp).$ By \Cref{thm:as}, this equals $\lambda((\pkpp, \row_\pkpp (\pk\cdot s)), \pkp)$, which by definition of $\lambda$ equals $T_\pkpp(\pk\cdot s, \pkp)$.

Now let $e = (\pkp\cdot\dup)\cdot e'$.
Since $\pk\cdot s\in S_\pkpp$, we know that $\pk\cdot s\cdot(\pkp\cdot\dup)\in S'_\pkp$. By closedness, there is a prefix $\xi\cdot s'\in S_\pkp$ such that $\row_\pkp(\pk\cdot s\cdot(\pkp\cdot\dup)) = \row_\pkp(\xi\cdot s')$. By \Cref{lem:tclosed}, we know $e'\in E_\pkp$.
Observing that $e'$ is smaller than $e$, we can reason:
\begin{align*}
\mathcal{H}(q_0(\pk), s\cdot e) &= \mathcal{H}(q_0(\pk), s\cdot (\pkp\cdot\dup)\cdot e') & e = (\pkp\cdot\dup)\cdot e'\\
&= \mathcal{H}(q_0(\xi), s'\cdot e') & \pk\cdot s\cdot(\pkp\cdot\dup) \equiv_\mathcal{H} \xi\cdot s', \Cref{lem:hmin}\\
&= T_\pkp(\xi\cdot s', e') & \text{Induction hypothesis}\\
&= T_\pkp(\pk\cdot s\cdot(\pkp\cdot\dup), e') & \row_\pkp(\pk\cdot s\cdot(\pkp\cdot\dup)) = \row_\pkp(\xi\cdot s')\\
&= T_\pkpp(\pk\cdot s, e) & e = (\pkp\cdot\dup)\cdot e'.
\end{align*}
\end{proof}

\repeatlemma{addstate}
\begin{proof}
Let $T'$ be the refined observation table giving rise to $\mathcal{H'}$. As proved by \Cref{thm:agree}, both $\mathcal{H}$ and $\mathcal{H'}$ correctly represent the observation data in $T$ and $T'$, respectively. As $c$ is a counterexample, $\mathcal H$ did not correctly classify it. As such, $c$ must not have been part of the observation data in $T$. Through the counterexample handling process, $c$ will be incorporated in the observation data of $T'$, and as such, $\mathcal H'$ will correctly classify it.

As both $\equiv_{\mathcal H}$ and $\equiv_{\mathcal H'}$ are refined by $\NKmne$, but $c$ does not have the same equivalence class in $\equiv_{\mathcal H}$ as it does in $\equiv_{\mathcal H'}$, and as $\mathcal H'$ still correctly classifies the observation data in $T$, it must be that $\equiv_{\mathcal H'}$ is then a strict refinement of $\equiv_{\mathcal H}$. As such, it must have at least one more equivalence class than $\equiv_{\mathcal H}$, and thus $\mathcal H'$ has at least one more state than $\mathcal H$.
\end{proof}

\subsection{$\SPP$ Learning Correctness}

\repeatlemma{select}
\begin{proof}
We will assume that any canonicalization of SPPs that is performed does not affect
the SPP semantics, so we will reason about the \emph{semantics} of a canonical SPP as though canonicalization does not occur (although the sizes and therefore the choice the algorithm makes will of course depend on canonicalization).
  We proceed by induction on the number of fields in $F$ (call this number $n$).
  For the base case, we consider $n=1$, i.e. $F=\{f_1\}$.
  Let $\Sem{e}(\pk, \pkp) = \ell$. We have 3 cases:
  \begin{itemize}
    \item Suppose $\pk.f_1 = v$ and $\pkp.f_1 = v$.
      Then $v\notin \keys(b) \cup \keys(m)$. Thus $\Sem{d}(\pk,\pkp) = \one$ iff $\ell =\one$, and as a result
      $\Sem{s}(\pk,\pkp)=\one$ iff $\ell=\one$.
    \item Suppose $\pk.f_1 = v$ but $\pkp.f_1 \neq v$.
      Then $v\notin \keys(b)$, but by construction we have $m[\pkp.f_1] = \ell$.
      and thus that $\Sem{s}(\pk,\pkp)=\one$ iff $\ell = \one$.
    \item Suppose $\pk.f_1 \neq v$.
      Then $v\in\keys(b)$, and $\ell = b[\pk.f_1][\pkp.f_1]$ by construction.
      Thus $\Sem{s}(\pk,\pkp)$ iff $\ell = \one$, as needed.
  \end{itemize}
  The inductive cases are similar in structure;
  we just need to be a bit more careful in dealing with them. Fix $n>1$.
  \begin{itemize}
    \item Suppose $\pk.f_n = v$ and $\pkp.f_n = v$.
      Then $v\notin \keys(b) \cup \keys(m)$. By induction, $\pkp\in\Sem{d}(\pk)$ iff $\ell =\one$, and as a result
      $\pkp\in\Sem{s}(\pk)$ iff $\ell=\one$.
    \item Suppose $\pk.f_n = v$ but $\pkp.f_n \neq v$.
      Then $v\notin\keys(b)$. By induction we have
      $\Sem{m[\pkp.f_n]}(\pk, \pkp)=\one$ iff $\ell=\one$, and thus that
      $\Sem{s}(\pk,\pkp)=\one$ iff $\ell = \one$.
    \item Suppose $\pk.f_n \neq v$.
      Then $\pk.f_n\in\keys(b)$, and by induction
$\Sem{b[\pk.f_n][\pkp.f_n]}(\pk,\pkp)=\one$ iff $\ell=\one$.
      Thus $\Sem{s}(\pk,\pkp)=\one$ iff $\ell = \one$, as needed.
  \end{itemize}
\end{proof}

\begin{rcorollary}{epp2spp}
\label{lem:epp2spp}
  Let $e$ be an $\EPP$, and let $h = \hypSPP\ e$. Then
  for all $(\pk,\pkp)$ for which $\Sem{e}(\pk,\pkp)$ is defined,
  then $\Sem{e}(\pk,\pkp) = \Sem{\hypSPP(e)}(\pk,\pkp)$.
\end{rcorollary}
\begin{proof}
  Observing that $\hypSPP$ returns the result of $\select$, the result follows from
  \Cref{lem:select}.
\end{proof}

\repeattheorem{sppterm}
\begin{proof}
Let the teacher have SPP $s$. The learner only adds examples to its EPP (i.e.
does not remove) and the labels are only assigned by the teacher, so the EPP
maintained is always consistent with $e$. Because of these facts and the fact
that $\Pk$ is finite, termination is trivial: the teacher must provide a
different counterexample each time, and once the learner has seen all of
$\Pk\times\Pk$, it must be correct on the next conjecture because of
\Cref{lem:epp2spp}.
\end{proof}

\subsection{$\SNKA$ Learning Correctness}

\repeatlemma{snkawelldef}
\begin{proof}
The finite set of states and start state are directly from the EA. The EA has only finitely many states because $P$ is finite and so has at most finitely many distinguishing suffixes for each packet space. The transition function $\delta$ and observation function $\varepsilon$ are well-defined by the definition of $\hypSPP$, and moreover the transition function meets the deterministic restriction because this is established directly by the construction of $\delta'$ in the definition of $\sym$.
\end{proof}

\repeatlemma{ea2snka}
\begin{proof}
Let $q,q' \in Q$ be arbitrary states.
We prove each part separately:
\begin{enumerate}[(a)]
\item We note that by the definition of $\ev$, we have $\Sem{\delta(q)(q')}(\pk,\pkp)=\one$ precisely when $q = \glob(\row_\pk(w))$ for some $w\in\Pref$, and $q' = \glob(\row_\pkp(w\cdot\pkp\cdot\dup))$.
Consistency of $P$ prevents there being a $w'$ for which $\row_\pk(w') = \row_\pk(w)$,
 and thus $\glob(\row_\pk(w')) = \glob(\row_\pk(w))$, but that
$\row_\pkp(w'\cdot\pkp\cdot\dup) \neq \row_\pkp(w\cdot\pkp\cdot\dup)$, which would mean
$\glob(\row_\pkp(w'\cdot\pkp\cdot\dup)) \neq \glob(\row_\pkp(w\cdot\pkp\cdot\dup))$ by restriction 2 of $\glob$ (\Cref{sec:glob}).
 As a result, there cannot be a $q'' \neq q'$ for which $\Sem{\delta(q)(q'')}(\pk,\pkp)=\one$. In fact, because of the addition of ``determinizing pairs'': for each $q''\neq q'$, we have
$\Sem{\delta(q)(q'')}(\pk,\pkp)=\zero$.
Applying \Cref{lem:epp2spp}, we have that
$\Sem{\hypSPP(\delta'(q)(q'))}(\pk,\pkp)=\one$, and
$\Sem{\hypSPP(\delta'(q)(q''))}(\pk,\pkp)=\zero$ for any $q''\neq q'$.
The last step in finalizing $\delta'$ is to take differences to determinize (end of \Cref{sec:ea2snka}). But because we have just shown $q$ to $q'$ is \emph{negative} for this transition, this will not affect our transition from $q$ to $q'$ on $(\pk,\pkp)$, and we have
$\Sem{\delta' (q) (q')} (\pk,\pkp) =\one$ as needed.
\item Follows immediately from \Cref{lem:epp2spp}.
\end{enumerate}
\end{proof}

\repeatlemma{excorrect}
\begin{proof}
By induction on $e$. For the base case $e = \pkp$ for $\pkp\in\Pk$. Then:
\begin{align*}
\mathcal{M}(\glob(\row_\pk(w), \pk\cdot e)) &\iff \mathcal{M}(\glob(\row_\pk(w), \pk\cdot \pkp))=\one &\text{ base case}\\
&\iff \Sem{\varepsilon'(s)}(\pk,\pkp)=\one &\text{definition of $\mathcal{M}$}\\
&\iff \Sem{\varepsilon(s)}(\pk,\pkp)=\one &\text{by \Cref{lem:ea2snka-preserve}\ref{lem:preserve-obs}}\\
&\iff T_\pk(w,\pkp) &\text{definition of $\ev$}
\end{align*}
For the inductive case, $e = \pkpp\cdot e'$, and let
$s_\pk = \glob(\row_\pk(w))$, and $s_\pkpp = \glob(\row_\pkpp(w\cdot\pkpp\cdot\dup))$. Then:
\begin{align*}
\mathcal{M}(s_\pk, \pkp\cdot \pkpp\cdot\dup\cdot e')
&\iff \mathcal{M}(s_\pkpp, \pkpp\cdot e')
            &\text{\Cref{lem:ea2snka-preserve}\ref{lem:preserve-trans} and }
            \Sem{\delta'(s_\pk)(s_\pkpp)}(\pk,\pkpp) =\one\\
&\iff T_\pkpp(w\cdot\pkpp\cdot\dup, e')=\one &\text{Induction hypothesis}\\
&\iff T_\pk(w, e) = \one &\text{definition of $e$}
\end{align*}
\end{proof}

\repeatcorollary{cexmonotone}
\begin{proof}
Let $\pk\cdot e = c$, and note that $\pk, e$ satisfy the requirements of \Cref{lem:ex-correct} by the definition of \Cref{fig:updatesymbolic}. So:
\begin{align*}
\mathcal{M}(s_0, \pk\cdot e)
&\iff \mathcal{M}(\glob(\row_\pk(\pk)), \pk\cdot e)=\one &\text{restriction 1 of $\glob$ (\Cref{sec:glob})}\\
&\iff T_\pk(\pk, e)=\one & \text{\Cref{lem:ex-correct}}\\
&\iff c\in \mathcal{L}
\end{align*}
\end{proof}

\repeattheorem{snkacorrect}
\begin{proof}
Correctness follows from termination because we only terminate upon a successful equivalence query.
There are 3 ways examples can be classified as wrong:
\begin{enumerate}[(1)]
\item The hypothesis merges two of the target's same-packet Myhill-Nerode classes into one state.
\item The hypothesis has a wrong transition between classes because of missing data in the table.
\item The hypothesis has a wrong observation of a class because of missing data in the table.
\end{enumerate}
Note that the first of these is the only one possible in the canonical learner (\Cref{fig:learncanonical}) because (2) is prevented by maintaining $S'_\pk$, and (3) is prevented because $\Pk\subseteq E_\pk$.
We have already shown each counterexample is classified correctly in subsequent hypotheses (\Cref{lem:cex-monotone}).
By \Cref{thm:mn}, we can only correct finitely many errors of type (1). Suppose we have identified $n$ Myhill-Nerode
classes in the current partial observation table (i.e., there are a total of $n$ distinct rows across all packet tables).
We can only make $n\cdot |\Pk|$ errors of type (2) and $n\cdot|\Pk|$ errors of type (3).
We conclude that we can receive only finitely many counterexamples before the hypothesis must be correct on the next conjecture.
\end{proof}

}{} 
\end{document}